\documentclass[journal]{new-aiaa}

\usepackage[utf8]{inputenc}
\usepackage{longtable}
\usepackage{graphicx}
\usepackage{xcolor}
\usepackage{amsmath}

\usepackage{amssymb}
\usepackage{amsthm}
\usepackage{enumerate}
\usepackage[version=4]{mhchem}
\usepackage[linesnumbered,ruled,vlined]{algorithm2e}
\usepackage{siunitx}
\usepackage{url}
\usepackage{doi}
\usepackage{multirow}
\usepackage{hyperref}
\usepackage{mathrsfs}
\usepackage{subdepth}
\usepackage{pgfplots}
\usepackage{longtable,tabularx}
\usepackage{appendix}
\usetikzlibrary{pgfplots.groupplots}
\usepackage[usestackEOL]{stackengine}
\usetikzlibrary{external}
\makeatletter
\newcommand*{\centerfloat}{%
  \parindent \z@
  \leftskip \z@ \@plus 1fil \@minus \textwidth
  \rightskip\leftskip
  \parfillskip \z@skip}
\makeatother

\makeatletter
\def\@firstoftwo@second#1#2#3#4#5{%
  \def\temp##1.##2\@nil{##2}%
   \temp#1\@nil}
\newcommand\sref[1]{%
   \expandafter\@setref\csname r@#1\endcsname\@firstoftwo@second{#1}%
}

\usepackage{xifthen}

\DeclareFontFamily{U}{BOONDOX-calo}{\skewchar\font=45 }
\DeclareFontShape{U}{BOONDOX-calo}{m}{n}{
  <-> s*[1.05] BOONDOX-r-calo}{}
\DeclareFontShape{U}{BOONDOX-calo}{b}{n}{
  <-> s*[1.05] BOONDOX-b-calo}{}
\DeclareMathAlphabet{\mathcalboondox}{U}{BOONDOX-calo}{m}{n}
\SetMathAlphabet{\mathcalboondox}{bold}{U}{BOONDOX-calo}{b}{n}
\DeclareMathAlphabet{\mathbcalboondox}{U}{BOONDOX-calo}{b}{n}

\newcommand{\F}{ \textsc{\textit{F}}}
\newcommand{\I}{ \textsc{\textit{I}}}
\newcommand{\R}{ \textsc{\textit{R}}}

\newcommand{\Qlq}{\mathcalboondox{Q}}
\newcommand{\Rlq}{\mathcalboondox{R}}
\newcommand{\Uscr}{\mathcalboondox{U}}
\newcommand{\Dscr}{\mathcalboondox{D}}

\renewcommand{\t}{^{\mbox{\tiny \sf T}}}

\newcommand{\inv}{^{- 1}}

\DeclareMathOperator{\tr}{tr}

\DeclareMathOperator{\erf}{erf}

\renewcommand{\d}{\mathrm{d}}
\let\Pr\relax
\DeclareMathOperator{\Pr}{\mathbb{P}}
\newcommand{\Var}{\mathrm{Var}}
\newcommand{\Cov}{\mathrm{Cov}}

\renewcommand{\Re}{\mathbb{R}}
\newcommand{\E}{\mathbb{E}}

\newcommand{\Ncal}{\mathcal{N}}

\renewcommand{\Re}{\mathbb{R}}

\DeclareSymbolFont{matha}{OML}{txmi}{m}{it}
\DeclareMathSymbol{\varv}{\mathord}{matha}{118} 
\DeclareMathSymbol{\varw}{\mathord}{matha}{119} 

\newtheorem{theorem}{Theorem}[section]
\newtheorem{problem}{Problem}[]
\newtheorem{remark}[theorem]{Remark}
\newtheorem{corollary}[theorem]{Corollary}
\newtheorem{lemma}[theorem]{Lemma}

\graphicspath{ {./img/} }

\title{Stochastic Entry Guidance}

\author{Jack Ridderhof%
\footnote{Graduate Student, School of Aerospace Engineering, Georgia Institute of Technology, Atlanta, GA, 30332, USA.}
and Panagiotis Tsiotras%
\footnote{David and Andrew Lewis Chair and Professor, School of Aerospace Engineering, and Institute for Robotics and Intelligent Machines, Georgia Institute of Technology, Atlanta, GA, 30332}}
\affil{Georgia Institute of Technology, Atlanta, GA, 30332}

\author{Breanna J. Johnson%
\footnote{Aerospace Engineer, Flight Mechanics and Trajectory Design Branch NASA JSC/EG5.}}
\affil{NASA Johnson Space Center, Houston, TX, 77058}

\begin{document}


    \maketitle
    \begin{abstract}
		In this paper, closed-loop entry guidance in a randomly perturbed atmosphere, using bank angle control, is posed as a stochastic optimal control problem.
		The entry trajectory, as well as the closed-loop controls, are both modeled as random processes with statistics determined by the entry dynamics, the entry guidance, and the probabilistic structure of altitude-dependent atmospheric density variations.
		The entry guidance, which is parameterized as a sequence of linear feedback gains, is designed to steer the probability distribution of the entry trajectories while satisfying bounds on the allowable control inputs and on the maximum allowable state errors.
		Numerical simulations of a Mars entry scenario demonstrate improved range targeting performance with approximately 50\% lower 1st and 99th percentile final range errors when using the developed stochastic guidance scheme as compared to the existing Apollo final phase algorithm.
    \end{abstract}

%

\section{Introduction}

\lettrine{T}{he}
successful Mars Science Laboratory (MSL) mission demonstrated, for the first time, guided entry at a planet other than Earth.
The vehicle flew with a trimmed angle of attack for positive lift with an $L/D$ of $0.24$, and a reaction control system banked the vehicle to modulate the vertical lift based on the predicted range-to-target \cite{Mendeck2014entry}.
Following parachute deployment and a powered descent phase, the Curiosity rover was deployed approximately 2 \si{km} from the target landing site \cite{Mendeck2014entry}.
Using the same entry, descent, and landing architecture, the Mars 2020 Perseverance rover successfully landed on February 18, 2021, approximately 2 \si{km} from the targeted touch-down point in Jezero Crater.
Future exploration missions will require further improved landing accuracy, perhaps on the order of meters, in order to preposition supplies or to study interesting geological phenomena \cite{Braun2006mars,Cianciolo2016human}.
Improvements to entry guidance performance will be crucial to increasing landing accuracy.

\begin{figure}
	\centering
	\includegraphics{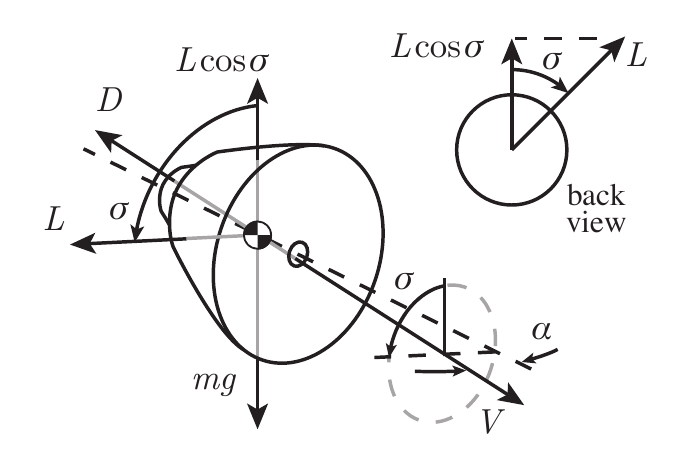}
	\caption{
		Bank angle control
		\label{fig:bank_angle_control}}
\end{figure}

The MSL entry guidance was derived from the Apollo final phase entry guidance, which includes a separate logic to independently set the magnitude of the bank angle (for range control) and the sign of the bank angle (for lateral control) \cite{Moseley1969apollo,Carman1998apollo}.
In this scheme, range control is achieved by scaling the vertical component of the lift vector, by means of setting the bank angle magnitude as shown in Figure~\ref{fig:bank_angle_control}, based on the predicted effect this change will have on the range flown.
Discrete bank-reversal events are triggered when the navigated crossrange error exceeds a threshold.
Underlying the Apollo final phase guidance is a mapping from \textit{constant} control input corrections to changes in the final downrange position.
In the case of the Apollo final phase range control, this mapping is approximated to first order (i.e., linearized) about a given reference trajectory, which, in turn, allows for the vertical lift correction to be written as a linear function of the current state deviation from this reference trajectory \cite{Moseley1969apollo,Carman1998apollo}.
The resulting onboard range control algorithm then only requires performing simple arithmetic after looking up the current set of feedback gains and nominal state values from a stored table.

The majority of modern proposed entry guidance algorithms follow the same basic principle as the Apollo final phase algorithm, in the sense that the control applied at any particular time should be those that, if held constant, will steer the vehicle to a target state.
However, due to improvements in onboard computational capabilities, guidance algorithms are no longer reliant on the linear approximation employed for Apollo \cite{Powell1998npc}.
Instead, the equations of motion can be numerically integrated onboard to obtain a trajectory resulting from a particular control input, and this process can be repeated for different candidate controls.
The requirement that the candidate control be constant, which is a necessary assumption in the derivation of the Apollo final phase feedback gains, is removed when using onboard numerical integration; however, the control function is still often low (one or two) dimensional as to support rapid and reliable onboard convergence \cite{Lu2014entry}.
This approach is referred to as numerical-predictor corrector (NPC) guidance \cite{Lu2014entry,Xue2010constrained,Putnam2010predguid}.

While in flight, the vehicle will most likely deviate from the planned trajectory due to external disturbances and parametric uncertainties.
Thus, the actual control inputs will not be equal to the predicted control inputs, and the actual trajectory will not be equal to the predicted trajectory.
The performance of an entry guidance algorithm is thus often measured statistically following repeated random Monte Carlo trials, which include variations to the atmosphere and initial vehicle states, in addition to a large number of additional randomized parameters.
Holding constant the statistics of the Monte Carlo inputs, and taking the number of trials to be a very large number, this procedure establishes a mapping from the \textit{algorithm} used for entry guidance to the \textit{statistics} of the closed-loop entry trajectory, which in this context is now a random process rather than a deterministic function of time.
One may then consider the entry guidance problem as finding a \textit{guidance algorithm}, or in certain cases a \textit{guidance law}, which results in desirable statistics of the closed-loop entry trajectory.
Unfortunately, it is difficult to directly use Monte Carlo for constrained guidance optimization.
For more insight into this problem, we thus turn to recent developments in stochastic control theory.

In recent years, the controlled evolution of the state statistics of linear systems has been extensively studied, and authors have proposed methods to solve for control laws which directly control the state probability distribution \cite{Chen2016b,Ridderhof2019nonlinear,Okamoto2018csl}.
For a general nonlinear stochastic system, one may equivalently consider either the random dynamics of sample paths or the deterministic evolution of the state probability distribution, which satisfies a partial differential equation (PDE) \cite{Brockett2012}.
In the case the system is linear with additive Brownian noise, then the PDE describing the state probability distribution can be decomposed into two independent ordinary differential equations (ODEs) for both the mean and the covariance of the state.
The nominal, or feedforward, control appears as the input to steer the evolution of the expected state, whereas the feedback gain appears as the input to steer the evolution of the state covariance.
Thus, one may solve for the nominal control to steer the mean of the state distribution, while also solving for the feedback gain to steer the covariance \cite{Brockett2012}.
From this perspective, stochastic control is concerned with controlling the \textit{deterministic} dynamics describing the system uncertainty (e.g., covariance) rather than controlling a collection of uncertain sample trajectories.

In this paper, we take a stochastic control approach to entry guidance following the aforementioned works on covariance control.
As for the Apollo final phase algorithm, the proposed stochastic entry guidance algorithm is assumed to consist of a reference trajectory, a linear feedback law, and table lookup; but the feedback gains in this proposed law are designed to steer the \textit{covariance} of the entry trajectory rather than steer a particular sample path.
By considering the closed-loop evolution of the state covariance, we are able to enforce constraints on the probability distributions of the closed-loop state and control, while minimizing the final range error variance.
Furthermore, we are able to quantify the effect on the final state covariance of terminating the entry trajectory as a function of the state, which is referred to as a state trigger, rather than simply a final time; in particular, the structure of the state trigger defines a transformation on the final state covariance.
The drift term (i.e., the deterministic dynamics) in the stochastic model is obtained by a linear approximation of the longitudinal entry dynamics evaluated about a given reference trajectory, similar to the Apollo final phase algorithm.
But for a stochastic treatment, it is also necessary to include the approximate effect of random atmospheric disturbances during entry, which is a nontrivial problem.

Monte Carlo studies indicate that for robotic class missions to Mars, dispersions in the atmospheric density and in the initial state of the vehicle are leading drivers of landing position uncertainty that must be controlled by closed-loop guidance, as navigation-derived uncertainty cannot be directly affected by guidance \cite{Dutta2017}.
While the initial vehicle state uncertainty may be simply modeled as a Gaussian random vector, the atmospheric density uncertainty is often modeled by the Global Reference Atmosphere Models (GRAMs), which include random variations in the density as a function of altitude \cite{Justus2002}.
More generally, an atmosphere model provides uncertainty as a function of position \cite{Tyler2008mesoscale}.
Thus, the density variations are a spatially-dependent random process, but, due to the vehicle motion through the atmosphere, the density variations at the vehicle position become a random process in time.
Based on this observation, we derive an expression for the diffusion coefficient (i.e., the noise intensity) as a function of the vehicle sink (descent) rate so that the entry trajectory can be expressed as a stochastic differential equation (SDE) driven by Brownian noise.

The proposed stochastic entry guidance depends on the given reference trajectory, the covariance of the initial vehicle states, and the intensity of density variations as a function of altitude.
From an operational perspective, the entry guidance feedback gains thus depend not only on the nominal trajectory, but also on the interplanetary delivery performance.
If an additional course correction that decreases the delivery uncertainty is performed, for example, then entry guidance performance could be improved by recomputing gains under the assumption of a smaller initial state uncertainty.

In contrast to the modern NPC approach to entry guidance, the proposed stochastic entry guidance is more closely related to the Apollo final phase guidance.
For instance, while NPC approaches remove the requirement of supplying a reference trajectory, the proposed method is based on linear perturbations about a given reference trajectory.
It follows that the applicability of the proposed method is restricted to correcting for relatively small deviations from this reference, whereas NPC-based entry guidance methods have shown to be robust to large state deviations \cite{Lu2014entry,Putnam2010predguid}.
On the other hand, NPC entry guidance requires more complex onboard calculations and does not always have theoretical guarantees on convergence.
The proposed guidance, as with the Apollo final phase algorithm, relies entirely on precomputed table values and does not require onboard trajectory recalculation.
In summary, the proposed method is not a general purpose guidance solution, but rather is an evolution of the Apollo final phase approach with improved range control performance.

The contributions of this paper are summarized as follows.
A novel stochastic process model of density variations is introduced, inspired by the GRAM density variation models, but which is given in an explicit form as an SDE.
The SDE representation of the density variation process is leveraged to derive a closed-form linear covariance model for atmospheric entry, which includes the effects of random density variations; in addition to guidance design, this linear covariance model may also be applied for approximate uncertainty quantification in preliminary trade studies \cite{Woffinden2019lincov,Carson2019splice}.
Lastly, we leverage this stochastic model to derive closed-loop entry guidance, while considering the effect of a state-dependent termination condition, which we demonstrate in a Monte Carlo simulation.

This paper is organized as follows.
In Section~\ref{sec:entry_as_sde}, a stochastic process model for atmospheric entry is developed, based on the proposed SDE model for the atmospheric density.
Next, in Section~\ref{sec:bam_range_control}, this stochastic entry model is used to derive a range control guidance using bank angle modulation.
A brief review of the Apollo final phase algorithm is included for completeness.
The effect of state triggers on the final state distribution, and hence on the range control law, is considered in Section~\ref{sec:state_triggers}.
Then, in Section~\ref{sec:numerical_examples}, the proposed stochastic entry guidance method is compared to the Apollo final phase guidance for a Mars entry mission based on MSL.
Finally, we conclude in Section~\ref{sec:conclusion}.

\section{Entry as a Stochastic Process}
\label{sec:entry_as_sde}
The motion of an entry vehicle in atmospheric gliding flight around a spherical, rotating planet is described in planet-relative coordinates by the system of equations \cite{Vinh1980,Johnson2020pterodactyl}
\begin{subequations} \label{eq:full_eom}
	\begin{equation}
		\dot{r} = V \sin \gamma
	\end{equation}
	\begin{equation}
		\dot{\theta} = \frac{V \cos \gamma \sin \psi}{r \cos \phi}
	\end{equation}
	\begin{equation}
		\dot{\phi} = \frac{V \cos \gamma \cos \psi}{r}
	\end{equation}
	\begin{equation}
		\dot{V} = -\frac{D}{m} - \frac{\mu \sin \gamma}{r^2} + \Omega ^ 2 r \cos \phi (\sin \gamma \cos \phi - \cos \gamma \sin \phi \cos \psi)
	\end{equation}
	\begin{equation}
		\dot{\gamma} = \frac{1}{V} \bigg( \frac{L \cos \sigma}{m} - \frac{\mu \cos \gamma}{r^2} + \frac{V^2 \cos \gamma}{r} + 2 \Omega V \cos \phi \sin \psi 
		+ \Omega ^ 2 r \cos \phi ( \cos \gamma \cos \phi + \sin \gamma \sin \phi \cos \psi) \bigg)
	\end{equation}
	\begin{equation}
		\dot{\psi} = \frac{1}{V} \bigg( \frac{L \sin \sigma}{m \cos \gamma} - \frac{V ^ 2 \cos \gamma \sin \psi \tan \phi}{r} 
			 + 2 \Omega V (\tan \gamma \cos \phi \cos \psi - \sin \phi) - \frac{\Omega ^ 2 r \sin \phi \cos \phi \sin \psi}{\cos \gamma}  \bigg)
	\end{equation}
\end{subequations}
where $r$ is the distance from the planet center to the vehicle, $\theta$ is the longitude, $\phi$ is the latitude, $V$ is the planet-relative velocity, $\gamma$ is the planet-relative flight path angle, and $\psi$ is the planet-relative heading azimuth.
The bank angle $\sigma$ is the out-of-plane angle between the lift vector and the local vertical, measured about the velocity vector in a right-hand sense; $\Omega$ is the planet rotation rate; $\mu$ is the planet gravitational parameter; $m$ is the vehicle mass, which is assumed to be constant;
and $L$ and $D$ are the lift and drag forces given by
\begin{align}
	L &= q A_{\text{ref}} C_L(\alpha) \\
	D &= q A_{\text{ref}} C_D(\alpha)
\end{align}
where $q = \rho V^2 / 2$ is the dynamic pressure in terms of the atmospheric density $\rho$, $A_{\text{ref}}$ is a reference area,
and $C_L$ and $C_D$ are given aerodynamic coefficients which depend on the angle of attack $\alpha$, which we assume to be a function of velocity given by the vehicle trim configuration.

At entry interface (EI), which is the time when the vehicle radius first equals the radius of the edge of the sensible atmosphere $r_{\mathrm{atm}}$, the vehicle state $z = z_0$ is assumed to be Gaussian distributed as
\begin{equation}
	z_0 = (r_0, \theta_0, \phi_0, V_0, \gamma_0, \psi_0) \sim \Ncal (\bar{z}_0, \Sigma_0)
\end{equation}
in terms of a known mean vector $\bar{z}_0$ and covariance matrix $\Sigma_0$.
Note that by this definition of the initial time, the initial radius variance is zero.
In practice, an initial state distribution with nonzero radius variance can be transformed into an alternate representation with zero radius variance by integrating Monte Carlo samples forward or backward to hit the EI radius.
An approximate method to perform this transformation is discussed in the context of state triggers in Section~\ref{sec:state_triggers}. 

The density $\rho$ is decomposed into a mean density $\bar{\rho}$ and a multiplicative density variation $\delta \rho$ as follows
\begin{equation}
	\rho = \bar{\rho} (1 + \delta \rho)
\end{equation}
The mean density is a function of the radial position and is given as the solution to the ordinary differential equation (ODE)
\begin{equation} \label{eq:hydro_equil}
	\frac{\d \bar{\rho}}{\d r}(r) = -\frac{\bar{\rho}(r)}{H(r)}
\end{equation}
where $H(r)$ is the scale height, and with a boundary condition $\bar{\rho}(r_p) = \bar{\rho}_{r_p}$ at the planet surface radius $r_p$.
The density variation, on the other hand, is assumed to be a stochastic process taking values as a function of the radial position.
For notational convenience, we define the \textit{sink distance} as
\begin{equation}
	s(r) = r_{\text{atm}} - r
\end{equation}
where $r_{\text{atm}}$ is the radius of the edge of the sensible atmosphere.
Since the sink distance and the radius have a unique correspondence, we will use $s$ and $r$ interchangeability when the context is clear.

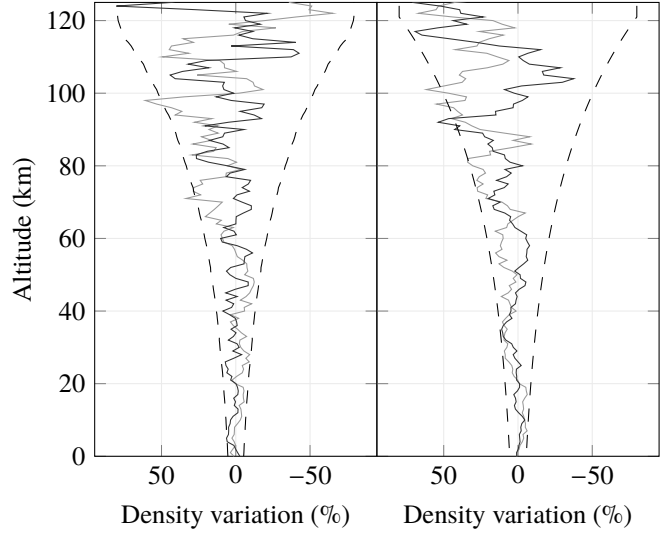
\begin{figure}
	\centering
	\tikzsetnextfilename{delrho_ou_compare}
	\begin{tikzpicture}
\begin{groupplot}[
    group style={
        group name=my plots,
        group size=2 by 1,
		x descriptions at=edge bottom,
		y descriptions at=edge left,
        vertical sep=0pt,
		horizontal sep = 0pt,
    },
	ylabel={Altitude (\si{km})},
	xlabel={Density variation (\%)},
	ymin=0, ymax=125,
	xmin=-95, xmax=+95,
	grid = major,
	grid style = {draw=black!8},
	ylabel near ticks,
	ylabel shift = -3 pt,
	width = 2.1in,
	height = 3in,
	x dir = reverse,
]

\def\numplot{20}
\def\stochasticName{Stochastic}
\def\colorScaleRange{40}
\def\colorRangeMin{20}


\nextgroupplot

\pgfplotstableread{data/mars_delrho_ou.txt}\loadedtable
\addplot [black!40, x filter/.expression={x*100},] table [y=h, x=s1] {\loadedtable};
\addplot [black!80, x filter/.expression={x*100},] table [y=h, x=s6] {\loadedtable};

\addplot [dashed,dash pattern=on 5pt,black!100, x filter/.expression={+x*200},] table [x index=1, y index=0] {data/mars_atmo_stddev_ou_mc.txt};
\addplot [dashed,dash pattern=on 5pt,black!100, x filter/.expression={-x*200},] table [x index=1, y index=0] {data/mars_atmo_stddev_ou_mc.txt};

%

\nextgroupplot

\pgfplotstableread{data/mars_delrho.txt}\loadedtable
		
\addplot [black!40, x filter/.expression={x*100},] table [y=h, x=s1] {\loadedtable};
\addplot [black!80, x filter/.expression={x*100},] table [y=h, x=s6] {\loadedtable};

\addplot [dashed,dash pattern=on 5pt,black!100, x filter/.expression={+x*200},] table [x index=1, y index=0] {data/mars_atmo_stddev.txt};
\addplot [dashed,dash pattern=on 5pt,black!100, x filter/.expression={-x*200},] table [x index=1, y index=0] {data/mars_atmo_stddev.txt};

%

\end{groupplot}

\end{tikzpicture}
	\caption{
		Comparison between OU process samples (left) with MarsGRAM density variation samples (right) with dashed $2\sigma$ limits
		\label{fig:delrho_ou_compare}}
\end{figure}

The density variation process is assumed to be a zero-mean Ornstein–Uhlenbeck (OU) process given by the stochastic differential equation (SDE)
\begin{equation} \label{eq:density_var_sde}
	\d \delta \rho(s) = - \lambda(s) \delta \rho (s) \d s + \sqrt{ \varphi(s) } \, \d w(s)
\end{equation}
where $w(s)$ is a standard Brownian motion, and where $\lambda(s)$ and $\varphi(s)$ are both non-negative functions which determine the structure of the atmospheric uncertainty.
The density variation at the zero sink position (the edge of the atmosphere) is normally distributed as
\begin{equation}
	\delta \rho(0) \sim \Ncal(0, \zeta_0)
\end{equation}
for some initial variance $\zeta_0 \geq 0$.
The density variation process (\ref{eq:density_var_sde}) is assumed to be an OU process since the OU process is both linear and Gaussian, and since the resulting model closely resembles the existing GRAM dispersion model \cite{Ridderhof2021entry}; in Section~\ref{sec:numerical_examples}, this model is successfully used to represent MarsGRAM density dispersions. 
For illustration, density variations samples from both an OU process and from MarsGRAM are shown in Figure~\ref{fig:delrho_ou_compare}; for this example, the coefficients $\lambda(s)$ and $\varphi(s)$ in (\ref{eq:density_var_sde}) were determined by the procedure described in the following remark.


\begin{remark} \label{remark:stationary_delrho_variance}
	Setting $\lambda(s) = 2 / H(s)$ results in density variations statistically similar to the GRAM \cite{Justus2002} density variation samples \cite{Ridderhof2021entry}.
	Furthermore, for constant $H(s) \equiv H_0$ and $\varphi(s) \equiv \varphi_0$, it can be shown that
	\begin{equation}
		\lim_{s \to \infty} \E (\delta \rho^2(s)) = \frac{\varphi_0 H_0}{4}
	\end{equation}
	Assuming rapid convergence of this limit, we set
	\begin{equation}
		\varphi(s) = \frac{4 \zeta_d(s)}{H(s)}
	\end{equation}
	so that the variance $\E (\delta \rho^2(s))$ approximates a desired altitude-dependent variance profile $\zeta_d(s)$.
\end{remark}

The density process may also be represented by an SDE.
Applying It\^{o}'s formula to the function $F(s, \delta \rho) = \bar{\rho}(s) (1 + \delta \rho)$, and substituting (\ref{eq:hydro_equil}) and (\ref{eq:density_var_sde}), we obtain
\begin{align}
	\d \rho (s) &= \frac{\partial F}{\d s}(s, \delta \rho(s))\, \d s + \frac{\partial F}{\partial \delta \rho}(s, \delta \rho(s)) \, \d \delta \rho(s) \\
	&= \underbrace{
			\bigg( \bigg( \frac{1}{H(s)} - \lambda(s) \bigg) \rho(s) + \lambda(s) \bar{\rho}(s) \bigg)
		}_{f_\rho(s, \rho(s))}
		\d s + \underbrace{
			\bar{\rho}(s) \sqrt{\varphi(s)}
		}_{g_\rho(s)} 
		\d w(s) \label{eq:density_sde_expanded}
\end{align}

It remains to combine the density process model with the vehicle dynamics in order to describe the motion of the vehicle in the random atmosphere.
First, we compactly rewrite the dynamics (\ref{eq:full_eom}) as
\begin{equation}
	\dot{z} = f_z (z, \rho, u),
\end{equation}
where $u = \sigma$ is the control input.
Let $s(t) = s(r(t))$ be the sink distance of the vehicle at time $t$, which has derivative $\dot{s}(t) = -V (t) \sin \gamma(t)$.
Applying a change of variables to the drift part of the density SDE (\ref{eq:density_sde_expanded}), we obtain the time integral
\begin{equation}
	\int_{s(t_0)}^{s(t)} f_\rho(y, \rho(y)) \, \d y =  \int_{t_0}^{t} f_\rho(s(\tau), \rho(s(\tau))) \dot{s}(z(\tau)) \, \d \tau 
\end{equation}
which, when substituted into the integral form of (\ref{eq:density_sde_expanded}), results in the expression for the density at the vehicle position:
\begin{equation} \label{eq:density_integral_with_change_of_var}
	\rho(s(t)) = \rho(s(t_0)) + \int_{t_0}^{t} f_\rho(s(\tau), \rho(s(\tau))) \dot{s}(z(\tau)) \, \d \tau + \int_{s(t_0)}^{s(t)} g_\rho(y) \, \d w(y)
\end{equation}
Finally, concatenating (\ref{eq:density_integral_with_change_of_var}) with the vehicle dynamics, we obtain an integral equation for the joint evolution of the density and the vehicle state as
\begin{equation} \label{eq:combined_nonlinear_vehicle_density_system}
	\begin{bmatrix}
		z(t) \\
		\rho(s(t))
	\end{bmatrix}
	= \begin{bmatrix}
			z_0 \\
			\rho(s(t_0))
		\end{bmatrix}
	+ \int_{t_0}^t \begin{bmatrix}
			f_z(z(\tau), \rho(s(\tau)), u(\tau)) \\
			f_\rho(s(\tau), \rho(s(\tau))) \dot{s}(z(\tau))
		\end{bmatrix}
		\d \tau
	+ \int_{s(t_0)}^{s(t)} \begin{bmatrix}
		0 \\ g_\rho(y)
	\end{bmatrix}  \d w(y)
\end{equation}
with the combined state at the initial time being distributed as
\begin{equation}
	\begin{bmatrix}
		z_0 \\
		\rho(s_0)
	\end{bmatrix} = x_0 \sim \Ncal (\bar{x}_0, P_0),
\end{equation}
where
\begin{equation}
	\bar{x}_0 = \begin{bmatrix}
		\bar{z}_0 \\ \bar{\rho}(s_0)
	\end{bmatrix}, \qquad P_0 =  \begin{bmatrix}
		\Sigma_0 & 0 \\
		0 & \Var (\rho(s_0))
	\end{bmatrix} 
\end{equation}
The initial vehicle state $z_0$ is uncorrelated with density since, by construction, the initial altitude is known exactly.

This equation represents the coupled nature of the density variations and the vehicle trajectory.
However, since the limits of the stochastic integral (\ref{eq:combined_nonlinear_vehicle_density_system}) depend on the state, it is difficult to solve this equation in its present form.


\subsection{Non-climbing Flight}

In general, the density $\rho(s(t))$ at the vehicle position is a random process taking values as a function of time, however, this process is not necessarily Markovian, nor can it be given as the solution to an SDE.
Indeed, since the density is defined as a function of altitude, if the vehicle enters a period of lofting, where it descends and then climbs, the density process will, in effect, reverse through previous values.

This poses both technical and practical issues.
If increments of the density process in time are not independent, then the covariance of the joint vehicle-density process cannot be described by an ODE, and methods from stochastic control theory developed for Brownian motion-driven random processes cannot be applied.
Practically, we risk an over-fitting effect since the true atmospheric density likely depends also on the longitude and latitude in addition to the altitude.
If the vehicle descends and then climbs, it may be unreasonable to take a past value of density to be exactly equal to the present value since the vehicle may have traveled hundreds of kilometers before reaching the previously experienced altitude.
A complete stochastic atmosphere model for such a situation should therefore include statistical correlations which depend on downrange distance traveled.

We leave this spatial modeling issue as a problem for future work \cite{Ridderhof2021gpcs}, and instead, in this paper we focus primarily on the case of direct entry when the vehicle is monotonically descending.
While this assumption excludes from consideration certain lofting or skip entry trajectories, our primary concern is range control for low $L/D$ vehicles at Mars, which traditionally do not include significant enough lofting to invalidate our assumptions.
Indeed, in Section~\ref{sec:numerical_examples}, the proposed range control is successfully applied to an MSL-like entry scenario.


Following this line of reasoning, and assuming that the vehicle is not climbing, we have
\begin{equation} \label{eq:non_climbing_assumption}
	\dot{s}(t) \geq 0 \;\; \text{for all} \;\; t \geq t_0
\end{equation}
Strict adherence to this assumption is largely technical; in practice, as will be shown in Section~\ref{sec:numerical_examples}, lofting flight may be reasonably captured by this model.
This assumption allows us to write the density process using a stochastic integral over a Brownian motion (i.e., as an SDE),  as is justified by the following lemma.

\begin{lemma}{(Time-Changed Brownian Motion)} \label{lem:brownian_time_change}
	Let $\tilde{w}(s)$ be a Brownian motion and let $s: [t_0, t_f] \to [0, s_f]$ be a smooth, non-decreasing function. 
	Then, there exists a Brownian motion $w(t)$ such that, almost surely, for all $t \in [t_0, t_f]$,
	\begin{equation} \label{eq:brownian_time_change_integral_form}
		\tilde{w}(s(t)) = \int_{t_0}^{t} \sqrt{ \dot{s}(z) } \, \d w(z)
	\end{equation}
\end{lemma}

Applying the time-change formula (\ref{eq:brownian_time_change_integral_form}) to the stochastic integral in (\ref{eq:combined_nonlinear_vehicle_density_system}), we obtain
\begin{equation} \label{eq:time_changed_sto_integral}
	\int_{s(t_0)}^{s(t)} \begin{bmatrix}
		0 \\ g_\rho(y)
	\end{bmatrix}  \d w(y) = \int_{t_0}^{t} \begin{bmatrix}
		0 \\ g_\rho(s(\tau))
	\end{bmatrix} \sqrt{\dot{s}(z(\tau))} \, \d w(\tau)
\end{equation}
Finally, we substitute the time-changed integral from (\ref{eq:time_changed_sto_integral}) into (\ref{eq:combined_nonlinear_vehicle_density_system}) to obtain the SDE
\begin{equation} \label{eq:combined_nonlinear_time_sde_general}
	\d x = f (x, u) \, \d t + g (x) \, \d w(t)
\end{equation}
with the coefficient functions
\begin{equation}
	f(x, u) = \begin{bmatrix}
		f_z(z, \rho, u) \\
		f_\rho(s, \rho) \dot{s}(z)
	\end{bmatrix}, \qquad g(x) = \begin{bmatrix}
			0 \\ g_\rho(s) \sqrt{\dot{s}(z)} 
		\end{bmatrix} 
\end{equation}
This SDE completely defines the stochastic process describing the vehicle entry trajectory under the assumption that the atmospheric density variations are as in (\ref{eq:density_var_sde}) and that the vehicle trajectory is monotonically descending.

%
%

%

\section{Bank-Angle Range Control}
\label{sec:bam_range_control}

In this section, we consider the problem of range control using bank angle feedback.
The vehicle is assumed to be in trimmed flight with positive lift (at zero bank) and zero sideslip, and control is effected by banking to tilt the lift vector.
Changes to the vertical component of the lift are made by setting the magnitude of the bank angle, while the direction (left or right) of the lift force follows from the sign of the bank angle.

Vertical lift, which depends on the cosine of the bank angle, affects the vehicle’s sink rate and thus its altitude.
Increasing the altitude decreases the density and thus decreases drag.
It follows that the velocity, and hence the range flown, may be affected by the vertical lift by means of setting the bank angle magnitude \cite{Moseley1969apollo}.
In other words, the vehicle can affect the range flown by increasing (or decreasing) the bank angle in order to fly through thicker (or thinner) atmosphere.
We therefore set the cosine of the bank angle to be the longitudinal control input:
\begin{equation}
	u_\ell = \cos \sigma
\end{equation}
Then, during flight, the bank angle command is given by
\begin{equation} \label{eq:bank_angle_cmd_cosinv}
	\sigma = b_{\mathrm{dir}} \cos \inv u_\ell
\end{equation}
where the bank direction $b_{\mathrm{dir}} \in \{-1, +1\}$ is set by a separate lateral control logic. 
For the purposes of range control, we use the simplified system for the longitudinal dynamics given by
\begin{subequations} \label{eq:entry_long_eom}
	\begin{equation}
		\dot{r} = V \sin \gamma
	\end{equation}
	\begin{equation}
		\dot{V} = - \frac{\rho V^2 A_{\text{ref}} C_D}{2 m} - \frac{\mu \sin \gamma}{r^2}
	\end{equation}
	\begin{equation}
		\dot{\gamma} = \frac{\rho V A_{\text{ref}} C_L}{2 m} \cos \sigma - \bigg( \frac{\mu}{r^2} - \frac{V^2}{r} \bigg) \frac{\cos \gamma}{V}
	\end{equation}
	\begin{equation}
		\dot{R} = V \cos \gamma
	\end{equation}
\end{subequations}
where $R$ is the downrange distance traveled.
In terms of the longitudinal vehicle state
\begin{equation}
	z_\ell = (r, V, \gamma, R)
\end{equation}
the longitudinal dynamics (\ref{eq:entry_long_eom}) are compactly rewritten as
\begin{equation} \label{eq:long_eom_short}
	\dot{z}_\ell = f_{z_\ell} (z_\ell, \rho, u_\ell)
\end{equation}
The range control problem is concerned with determining the longitudinal control inputs $u_\ell$ that steer the evolution of the longitudinal dynamics to reach the target conditions.
In the following subsections, the control will be parameterized as a linear function of the state, and thus the range control problem will reduce to identifying the state feedback gains.
First, the longitudinal dynamics (\ref{eq:long_eom_short}) are approximated by a linear discrete-time system with an associated finite sequence of feedback gains.

\subsection{Linear Discrete-Time Model}

We assume that a reference trajectory $\hat{z}_\ell(t)$, with corresponding nominal density $\hat{\rho}(t) = \bar{\rho}(\hat{r}(t))$, and reference bank profile $\hat{\sigma}(t)$ are provided, as is done in the Apollo direct entry method \cite{Moseley1969apollo,Carman1998apollo}.
See Ref.~\cite{Mendeck2014entry} for a discussion of the reference bank profile design process.
The deviation of the combined vehicle and density state from this reference is denoted by
\begin{equation}
	\tilde{x}_\ell(t) = \begin{bmatrix}
		z_\ell(t) - \hat{z}_\ell(t) \\
		\rho(t) - \hat{\rho}(t)
	\end{bmatrix}
\end{equation}
and the corrective control input is denoted by $\tilde{u}_\ell(t)$.
The closed-loop bank angle magnitude is given by the sum of the nominal part and the corrective part:
\begin{equation} \label{eq:long_control_decomp}
	u_\ell = \cos \hat{\sigma} + \tilde{u}_\ell
\end{equation}

The combined longitudinal vehicle dynamics and density process are linearized about this reference to obtain a linear stochastic system for the state deviation dynamics, given by
\begin{equation} \label{eq:long_linear_continuous}
	\d \tilde{x}_\ell(t) = (A_\ell (t) \tilde{x}_\ell(t) + B_\ell (t) \tilde{u}_\ell(t)) \d t + G_\ell(t) \d w(t)
\end{equation}
where the coefficient matrices
\begin{equation} \label{eq:long_sys_mats}
	A_\ell (t) = \frac{\partial}{\partial z_\ell} \begin{bmatrix}
		f_{z_\ell} \\ f_\rho \dot{s}
	\end{bmatrix}, \quad B_\ell (t) = \frac{\partial}{\partial u_\ell} \begin{bmatrix}
		f_{z_\ell} \\ f_\rho \dot{s}
	\end{bmatrix}, \quad  G_\ell(t) = \begin{bmatrix}
			0_{4 \times 1} \\ g_\rho \sqrt{\dot{s}}
		\end{bmatrix}
\end{equation}
are evaluated along the reference trajectory, and where, as before, $w(t)$ is a one-dimensional standard Brownian motion.

The corrective control input is assumed to be  constant on the subintervals of a partition $\mathscr{P} = (t^p_k)_{k = 0}^{N_p}$ of the interval $[t_0, t_f]$ given by
\begin{equation}
	t_0 = t_0^p < t_1^p < \dots < t_{N_p}^p = t_f
\end{equation}
In the derivation of the Apollo final phase entry guidance, in contrast, the corrective control at any time is assumed to be constant for the remainder of the flight \cite{Moseley1969apollo,Carman1998apollo}.
However, the bank angle corrections realized in flight are \textit{not} constant in the Apollo implementation, whereas the control corrections in the present derivation \textit{will} in fact be constant on the subintervals of the partition $\mathscr{P}$.
Maintaining this more representative model of the closed-loop bank angle corrections will allow us to leverage the degrees of freedom given by the number of subintervals in $\mathscr{P}$ to enforce constraints on the entry trajectory and on the closed-loop bank angle.

Since, for any step $k \in \{0, \dots, N_p - 1\}$, the control correction $\tilde{u}_\ell(t)$ is constant on the interval $[t^p_k, t^p_{k + 1}]$, we can integrate the continuous-time system (\ref{eq:long_linear_continuous}) from $t^p_k$ to $t^p_{k + 1}$ and obtain the discrete-time system
\begin{equation}
	\tilde{x}_{\ell, k + 1} = A_{\ell, k} \tilde{x}_{\ell, k} + B_{\ell, k} \tilde{u}_{\ell, k} + G_{\ell, k} w_k
\end{equation}
where
\begin{equation}
	\tilde{x}_{\ell, k} = \tilde{x}_\ell(t^p_k), \quad \tilde{u}_{\ell, k} = \tilde{u}_\ell(t^p_k)
\end{equation}
\begin{equation}
	A_{\ell, k} = \Phi_\ell(t^p_{k + 1}, t^p_k), \quad B_{\ell, k} = \int_{t^p_k}^{t^p_{k + 1}} \Phi_\ell(t^p_{k + 1}, \tau) B_\ell(\tau) \, \d \tau
\end{equation}
and where $\Phi_\ell(t,\tau)$ is the state transition matrix corresponding to the state matrix $A_\ell(t)$, which is the solution to the ODE
\begin{equation} \label{eq:long_stm}
	\frac{\partial}{\partial t} \Phi_\ell(t, t_0) = A_\ell(t) \Phi_\ell(t, t_0), \quad \Phi_\ell(t_0, t_0) = I
\end{equation}
and which satisfies the property that $\Phi_\ell(t, \tau) = \Phi_\ell(t, t_0) \Phi_\ell \inv (\tau, t_0)$.
The noise increments $w_k$ are independent standard Gaussian random vectors, and the matrices $G_{\ell, k}$ are set so that the random vector $G_{\ell, k} w_k$ has covariance
\begin{equation} \label{eq:long_discrete_noise_covariance}
	\int_{t^p_k}^{t^p_{k + 1}} \Phi_\ell(t^p_{k + 1}, \tau) G_\ell(\tau) G\t_\ell(\tau) \Phi_\ell \t (t^p_{k + 1}, \tau) \, \d \tau
\end{equation}
It follows that $G_{\ell, k}$ can be any matrix such that $G_{\ell, k} G_{\ell, k} \t$ equals the integral (\ref{eq:long_discrete_noise_covariance}), and, accordingly, the matrix product $G_{\ell, k} G_{\ell, k} \t$ (not just the matrix $G_{\ell, k}$) affects the evolution of the state covariance.

The control corrections are parameterized as linear functions of the state deviation, which results in the linear feedback law 
\begin{equation} \label{eq:long_feedback_law}
	\tilde{u}_{\ell, k} = K_{\ell, k} \tilde{x}_{\ell, k}
\end{equation}
in terms of the yet to-be-determined feedback gain matrices $K_{\ell, k} \in \Re ^{1 \times 5}$ for $k = 0, \dots, N_p - 1$.
Provided this feedback law, the covariance of the combined vehicle-density state 
\begin{equation}
	P_{\ell, k} = \E (\tilde{x}_{\ell, k} \tilde{x}_{\ell, k} \t)
\end{equation}
is obtained as the solution to the difference equation
\begin{equation} \label{eq:long_cov_dynamics}
	P_{\ell, k + 1} = (A_{\ell, k} + B_{\ell, k} K_{\ell, k}) P_{\ell, k} (A_{\ell, k} + B_{\ell, k} K_{\ell, k}) \t + G_{\ell, k} G_{\ell, k} \t
\end{equation}
with the initial condition
\begin{equation}
	P_{\ell,0} = \begin{bmatrix}
		\Sigma_{\ell, 0} & 0 \\
		0 & \Var\,( \rho(t_0) )
	\end{bmatrix}
\end{equation}
where the value of the initial longitudinal state covariance matrix $\Sigma_{\ell, 0}$ follows from the full state covariance matrix $\Sigma_0$.
There is initially no correlation between the vehicle states and the density since the initial vehicle radius is by definition fixed to be equal to the radius of the edge of atmosphere.
The variance of the density at the initial altitude is a function of the density variation variance $\zeta(s) = \Var\,(\delta \rho(s))$ given by
\begin{equation}
	\Var\,( \rho(t_0) ) = \zeta(s(t_0)) \bar{\rho} ^ 2(s(t_0))
\end{equation}
The closed-loop control $u_\ell(t)$ is also Gaussian distributed with mean $\cos \hat{\sigma}(t)$ and variance
\begin{equation}
	\Var \, (u_\ell(t)) = K_{\ell, k} P_{\ell, k} K_{\ell, k} \t
\end{equation}
for all times in the interval $[t^p_k, t^p_{k + 1}]$.
The distribution of the closed-loop control therefore depends on the state covariance and the feedback gain.

\subsection{Apollo Range Control}

Before presenting the stochastic approach to range control, and for the sake of completeness, we briefly review the derivation of the Apollo final phase range control as presented in Refs.~\cite{Moseley1969apollo,Carman1998apollo}.
As before, a nominal bank angle profile and the corresponding longitudinal trajectory are provided.
The dynamics are linearized about this reference to obtain a linear time-varying system as in (\ref{eq:long_linear_continuous}), except that the stochastic term is neglected.
Without the stochastic term, there is no longer a reason to include density as a state, and as such, the Apollo direct entry derivation takes the state to be equal to the vehicle state.
However, in an effort to keep the notation consistent, we will use the system matrices $A_\ell(t)$ and $B_\ell(t)$ as in (\ref{eq:long_sys_mats}) and the state transition matrix $\Phi_\ell(t, \tau)$ as in (\ref{eq:long_stm}).

At any time $t$, the corrective control input $\tilde{u}_{\ell, \mathrm{ap}}(t)$ is assumed to remain constant for the remainder of the flight.
The state deviation at the final time is therefore given by
\begin{equation}
	\tilde{x}_{\ell}(t_f) = \Phi_\ell(t_f, t) \tilde{x}_\ell (t) + \bigg( \int_{t}^{t_f} \Phi_\ell(t_f, \tau) B_\ell(\tau) \, \d \tau \bigg) \tilde{u}_{\ell, \mathrm{ap}}(t)
\end{equation}
Define the final total range error $\Delta R_f$ to be a linear function of the final state, given by
\begin{equation} \label{eq:apollo_total_range_error}
	\Delta R_f = \vartheta_f \t  \tilde{x}_\ell(t_f)
\end{equation}
for some influence-weighting vector $\vartheta_f$.
Note that $\Delta R_f$ is not necessarily equal to $\tilde{R}(t_f)$, since errors in the altitude may also contribute to the range error following parachute deployment \cite{Moseley1969apollo, Carman1998apollo}; the relationship between $\Delta R_f$ and $\tilde{R}(t_f)$ will be described in Section~\ref{sec:state_triggers}.
It follows that the total range error can be computed as a linear function of the state and control at any time along the trajectory in terms of the \textit{adjoint} functions $\vartheta(t)$ and $\vartheta_u(t)$ by
\begin{equation} \label{eq:final_total_range_error_adjoints}
	\Delta R_f = \vartheta \t (t) \tilde{x}_\ell (t) + \vartheta_u (t) \tilde{u}_{\ell, \mathrm{ap}}(t)
\end{equation}
The adjoint system is obtained as the solution to the backwards ODE \cite{BrysonHo}
\begin{equation}
	\frac{\d}{\d t} \begin{bmatrix}
		\vartheta(t) \\ \vartheta_u(t)
	\end{bmatrix} = -\begin{bmatrix}
		A_\ell \t (t) & 0 \\
		B_\ell \t (t) & 0
	\end{bmatrix} \begin{bmatrix}
		\vartheta(t) \\ \vartheta_u(t)
	\end{bmatrix}, \qquad \begin{bmatrix}
		\vartheta(t_f) \\ \vartheta_u(t_f)
	\end{bmatrix} = \begin{bmatrix}
		\vartheta_f \\ 0
	\end{bmatrix}
\end{equation}
Finally, the feedback law
\begin{equation} \label{eq:apollo_feedback_law}
	\tilde{u}_{\ell, \mathrm{ap}}(t) = K_{\ell, \mathrm{ap}}(t) \tilde{x}_\ell (t), \quad  K_{\ell, \mathrm{ap}}(t) = - K_{\mathrm{oc}} \frac{\vartheta \t (t)}{\vartheta_u(t)}
\end{equation}
is obtained by setting $\Delta R_f = 0$ in (\ref{eq:final_total_range_error_adjoints}) and solving for the control, and where $K_{\mathrm{oc}}$, which is referred to as the overcontrol gain, is a user-defined parameter which may be tuned to improve performance.
In practice, the feedback law (\ref{eq:apollo_feedback_law}) is rewritten in terms of range, drag, and climb rate feedback.
In addition, the feedback gain and the nominal trajectory are recast as functions of velocity with the nominal trajectory serving as the mapping between time and velocity.


\subsection{Stochastic Range Control}

In the absence of the stochastic forcing term in the linearized system (\ref{eq:long_linear_continuous}), the predicted range traveled by the vehicle would be controlled by the selection of a particular bank angle correction to be made during flight.
Accordingly, from a deterministic perspective, the central model for range control is the mapping defined by the linearized system (\ref{eq:long_linear_continuous}) of bank angle corrections to a predicted final range error; in the derivation of the Apollo final phase entry guidance this mapping is defined via the adjoint state, which determines the feedback gains.

When including the stochastic term, in contrast, there is no longer a unique mapping from bank angle corrections to range-error predictions, since, for any particular bank angle, the range flown is random.
The central model for stochastic range control instead becomes the state covariance evolution equation (\ref{eq:long_cov_dynamics}), which defines a mapping from bank angle \textit{feedback gains} to the predicted final range error \textit{variance}.
The stochastic range control problem is thus to solve for the feedback gains $K_{\ell, k}$ that minimize a function of final state covariance subject to constraints on the probability distribution of the final state and of the control inputs.

Constraints on the final state error are given in the form 
\begin{equation} \label{eq:long_final_state_constraints}
	\Pr ( \vert d_i \t \tilde{x}_\ell(t_f) \vert \leq \Delta^x_i ) \geq 1 - p^x_i
\end{equation}
for $i = 1, \dots, N_x$.
The vectors $d_i \in \Re ^ 5$ and scalars $\Delta^x_i$ define regions which the state error must lie within, and $p^x_i \in (0, 1]$ is the maximum probability with which this constraint may be violated.
For example, constraining the final altitude error to be less than 5 \si{km} with 99\% probability translates to the values $d_i = (1, 0, 0, 0, 0)$, $\Delta^x_i = 5$ \si{km}, and $p^x_i = 0.01$.

Assuming the state to be Gaussian distributed, the linear feedback law (\ref{eq:long_feedback_law}) implies that the control corrections are also Gaussian distributed.
However, the range of allowable closed-loop bank angles may be limited due to concerns of lateral control authority \cite{Mendeck2014entry}.
In flight, of course, commanded control corrections which are outside of a user-defined interval will be saturated, but including a saturation function in the law (\ref{eq:long_feedback_law}) would invalidate the linear structure.
Instead, the \textit{probability} that closed-loop bank angle commands saturate will be constrained, and through this constraint we establish a balance between the intensity of planned corrective controls and the allowable range of bank angles; in other words, this constraint establishes for the feedback gains a dependence on the allowable control inputs.
To this end, longitudinal control input deviations (i.e., deviations of the cosine of the bank angle) from the nominal value are constrained at each control decision time $t^p_k$ by
\begin{equation} \label{eq:long_control_constraints}
	\Pr ( \vert \tilde{u}_{\ell, k} \vert \leq \Delta^u_k ) \geq 1 - p^u_k
\end{equation}
for $k = 0, \dots, N_p - 1$, where $p^u_k \in (0, 1]$ is a user-defined maximum allowable probability that the commanded control is outside the allowable range.
This constraint assumes a symmetric constrained region $[-\Delta^u_k, +\Delta^u_k]$ about the nominal control value, since the control is Gaussian distributed and the selection of the feedback gains only affects the variance of $\tilde{u}_{\ell, k}$.
The closed-loop longitudinal control can be constrained to lie in the interval $[u_{\ell, k, \mathrm{min}}, u_{\ell, k, \mathrm{max}}]$ with probability at least $1 - p^u_k$ by taking $\Delta^u_k = \min \{ u_{\ell, k, \mathrm{max}} - \cos \hat{\sigma}(t^p_k), \cos \hat{\sigma}(t^p_k) - u_{\ell, k, \mathrm{min}} \}$.

The range control cost is a function of the sequence of feedback gains $K = (K_{\ell, k})$ given by
\begin{equation} \label{eq:long_range_cost}
	J_\ell(K) = \Var \, ( a_f \t \tilde{x}_\ell(t_f) ) + \sum_{k = 0}^{N_p - 1} \mathcal{R}_k \Var \, (\tilde{u}_{\ell, k})
\end{equation}
where the vector $a_f$ weights the coordinates of the final state error, and $(\mathcal{R}_k)$ is a sequence of non-negative control cost weights.
The vector $a_f$ may be defined, for instance, so that $a_f \t \tilde{x}_\ell(t_f) = \tilde{R}(t_f)$, and the control weights $(\mathcal{R}_k)$ can be set to be a small number so that bank angle corrections will only be commanded if there is a meaningful reduction in the final state covariance.
The stochastic range control problem is summarized as follows.

\begin{problem} \label{prob:stochastic_range_control}
	Find the sequence of feedback gains $K = (K_{\ell, k})$ that minimize the cost (\ref{eq:long_range_cost}), subject to the covariance dynamics (\ref{eq:long_cov_dynamics}), the control constraints (\ref{eq:long_control_constraints}), and the final state constraints (\ref{eq:long_final_state_constraints}).
\end{problem}

In general, Problem~\ref{prob:stochastic_range_control} is a discrete-time, chance-constrained, linear-quadratic-Gaussian (LQG) stochastic optimal control problem.
In the following, we establish a connection between Problem~\ref{prob:stochastic_range_control} and the \textit{unconstrained} LQG problem, which has a known closed-form solution.
This connection allows us to obtain the solution to Problem~\ref{prob:stochastic_range_control} by selecting proper LQG weights.
While this class of problems can be solved via convex programming \cite{Chen2016b,Ridderhof2019nonlinear}, we have observed that for the range control problem, the solution method via LQG is numerically better-behaved, is easier to implement, and is more intuitive (i.e., not a ``black-box'').

\subsubsection{Solution using Unconstrained Linear Quadratic Control}

The unconstrained LQG problem is summarized as follows.
\begin{problem} \label{prob:lq}
	Find the sequence of feedback gains $K = (K_{\ell, k})$ that minimize the cost
	\begin{equation} \label{eq:std_lqg_cost}
		J_{LQ}(K; \Qlq, \Rlq) = \tr \Qlq \Cov( \tilde{x}_\ell(t_f) )+ \sum_{k = 0}^{N_p - 1} \Rlq_k \Var \, (\tilde{u}_{\ell, k})
	\end{equation}
	for a given final state weight matrix $\Qlq \geq 0$ and sequence of positive, scalar control weights $\Rlq = (\Rlq_k)$, subject to the covariance dynamics (\ref{eq:long_cov_dynamics}).
\end{problem}
The solution to Problem~\ref{prob:lq} is obtained by the backwards equations \cite{BrysonHo}:
\begin{subequations} \label{eq:lq_solution}
\begin{align}
	K_{\ell, k} &= - (B_{\ell, k} \t S_{\ell, k + 1} B_{\ell, k} + \Rlq_k) B_{\ell, k} \t S_{\ell, k + 1} A_{\ell, k}, \\
	S_{\ell, k} &= A_{\ell, k} \t S_{\ell, k + 1} A_{\ell, k} - K_{\ell, k} \t (\Rlq_k + B_{\ell, k} \t S_{\ell, k + 1} B_{\ell, k}) K_{\ell, k}, \\
	S_{N_p} &= \Qlq
\end{align}
\end{subequations}
Comparing Problems~\ref{prob:stochastic_range_control} and \ref{prob:lq}, we observe that the LQG cost function (\ref{eq:std_lqg_cost}) appears to be the Lagrangian for Problem~\ref{prob:stochastic_range_control} with the outer product $a_f a_f \t$ corresponding to the weight $\Qlq$, which would imply that, for particular values of $\Qlq$ and  $\Rlq$ corresponding to Lagrange multipliers, these problems have the same solution.
The following theorem shows that this indeed the case.

\begin{theorem} \label{thm:lq_equiv}
	For particular values of the weights $\Qlq$ and $\Rlq$, the solution of Problem~\ref{prob:lq} solves Problem~\ref{prob:stochastic_range_control}.
\end{theorem}

\begin{corollary} \label{cor:solve_via_QR}
	The solution to Problem~\ref{prob:stochastic_range_control} may be obtained by optimizing the objective (\ref{eq:long_range_cost}) over $\Qlq$ and $\Rlq$, with the resulting gain $K$ given by (\ref{eq:lq_solution}), subject to the constraints (\ref{eq:long_control_constraints}) and (\ref{eq:long_final_state_constraints}).
\end{corollary}

As a consequence of this theorem and its corollary, the optimization variable $K$ for Problem~\ref{prob:stochastic_range_control}, which is $5 \times N_p$ dimensional, may be replaced with the pair $\Qlq$ and $\Rlq$, which is only $N_x(N_x + 1) / 2 + N_p$ dimensional (since $\Qlq$ is symmetric).
Furthermore, for each guess of the new decision variables, the resulting control law enjoys the properties of optimal LQG solutions, such as smooth feedback gains and negligible corrections when feedback is not beneficial.
For example, at low dynamic pressure, changes to the bank angle have a negligible effect on the dynamics.
Thus, through an optimization which penalizes control actions against the trajectory response, bank angle corrections will not be commanded when the dynamic pressure is low.

\subsection{Lateral Control}

The lateral control logic sets the bank direction $b_{\mathrm{dir}}$ as a function of the crossrange error $\varepsilon$, which, in turn, is a function of the current vehicle state, the target position, and the estimated final time $t_{f, \mathrm{est}}$.
See Figure~\ref{fig:rangedef} and Appendix~\sref{sec:targeting_calculations} for details on the downrange and crossrange calculations.
In this paper, we consider a simple deadband for lateral control, as in Refs.~\cite{Moseley1969apollo,Mendeck2014entry}, which depends on a user-specified maximum allowable crossrange $\varepsilon_{\mathrm{max}}(t, x)$.
The deadband should therefore approximately contain the crossrange error to be bounded as
\begin{equation}
	-\varepsilon_{\mathrm{max}}(t, x) \leq \varepsilon \leq +\varepsilon_{\mathrm{max}}(t, x)
\end{equation}
At discrete update times, a new bank direction $b_{\mathrm{dir}}^+$ is determined as a function of the current bank direction, the crossrange error, and the current deadband value given by
\begin{equation} \label{eq:deadband_logic}
	b_{\mathrm{dir}}^+(t, x, b_{\mathrm{dir}}) = \begin{cases}
		+1 & \text{if} \; b_{\mathrm{dir}} = -1 \; \text{and} \; \varepsilon < -\varepsilon_{\mathrm{max}}(t, x) \\
		-1 & \text{if} \; b_{\mathrm{dir}} = +1 \; \text{and} \; \varepsilon > + \varepsilon_{\mathrm{max}}(t, x) \\
		b_{\mathrm{dir}} & \text{otherwise}
	\end{cases}
\end{equation}

\subsection{Heading Alignment}

\begin{figure}
	\centering
	\includegraphics{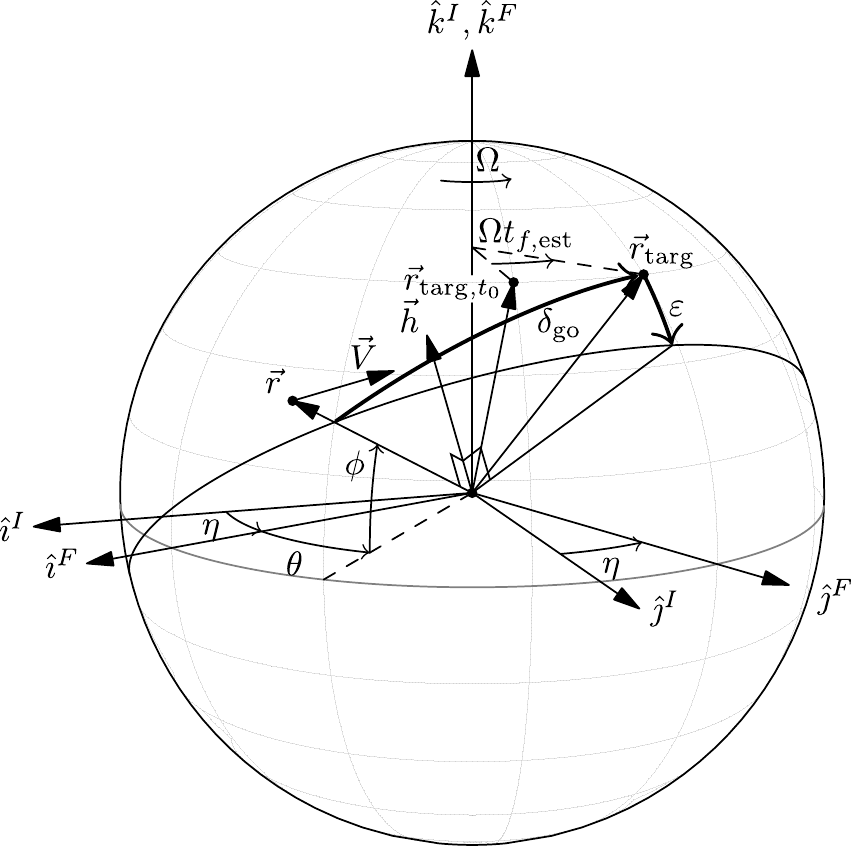}
	\caption{
		Downrange and crossrange definitions (frame definitions are given in Appendix~\protect\sref{sec:targeting_calculations})
		\label{fig:rangedef}}
\end{figure}

As the vehicle slows down and approaches the target, bank angle modulation becomes less effective at controlling the downrange position \cite{Mendeck2014entry}.
Therefore, when the vehicle passes some user-defined threshold, such as a minimum velocity, the guidance mode switches from range control to heading alignment.
During the heading alignment segment in Apollo final phase guidance, the bank angle commands are given by the proportional feedback law
\begin{equation} \label{eq:ha_law}
	\sigma = K_{\mathrm{ha}} \tan \inv \bigg( \frac{\delta_{\mathrm{go}}}{\varepsilon} \bigg)
\end{equation}
where $K_{\mathrm{ha}}$ is a user-defined heading alignment gain, and $\delta_{\mathrm{go}}$ and $\varepsilon$ are the downrange-to-go and crossrange angles \cite{Mendeck2014entry}, which are shown in Figure~\ref{fig:rangedef}.
The heading alignment law (\ref{eq:ha_law}) is adopted for the proposed stochastic guidance, as the focus of this paper is the range control phase.

\section{Range Control with Final State Triggers}
\label{sec:state_triggers}

In the previous section, the trajectory was considered over a given time interval with a fixed final time.
The guidance objective was therefore to steer the state --- or, in the stochastic case, the state covariance --- to a desired value at this fixed final time.
In practice, however, entry trajectories are rarely terminated at a final time.
Rather, trajectories are terminated by a condition on the vehicle state.
The MSL entry guidance, for example, ended (by initializing the parachute deployment sequence) when the navigated vehicle velocity passed below a threshold \cite{Mendeck2014entry}.
More generally, any condition on the vehicle state, which we refer to as a state trigger, can mark the end of guided entry.
Selection of the state trigger criteria can have a significant effect on the final state statistics, as has been demonstrated in Monte Carlo simulations for the Mars 2020 mission, in which switching from a velocity to a range trigger has been shown to significantly improve landing accuracy \cite{Dutta2017}.

We are thus motivated to study the effect of the state trigger on the final state statistics.
The following analysis is a generalization of the derivation of the final state weight developed in Ref.~\cite{Moseley1969apollo}.
Formally, we define a state trigger as a hyperplane in the state space described by the vector $\nu$ and the scalar $\beta$.
The time that the vehicle hits the state trigger is the stopping time
\begin{equation} \label{eq:stopping_time_def}
	T = \inf \{ t \geq t_0 : \nu \t x(t) \leq \beta \}
\end{equation}
We assume that the nominal trajectory terminates at this stopping time, and thus
\begin{equation} \label{eq:stopping_time_mean_condition}
	\nu \t \bar{x}(t_f) = \beta
\end{equation}

Approximating to first order the state drift from the final time $t_f$ to the stopping time $T$, we obtain
\begin{equation} \label{eq:approximate_state_at_stopping_time}
	x(T) \approx x(t_f) + (T - t_f) \hat{f}_f 
\end{equation}
where $\hat{f}_f = f(\hat{x}(t_f), \hat{u}(t_f))$ is the nominal state derivative at the final time.
Note that we neglect the diffusion term, since the effect of density variations during this short period is insignificant.
Rearranging the terms in (\ref{eq:approximate_state_at_stopping_time}), and using the condition (\ref{eq:stopping_time_mean_condition}), we obtain
\begin{equation}
	\nu \t \big( x(T) - x(t_f) \big) = \nu \t \big( \bar{x}(t_f) - x(t_f) \big) \approx (T - t_f) \nu \t \hat{f}_f 
\end{equation}
which after simplification yields
\begin{equation} \label{eq:stopping_time_approx}
	T \approx t_f - \bigg( \frac{\nu \t}{\nu \t \hat{f}_f} \bigg) \tilde{x}(t_f)
\end{equation}
The stopping time $T$ is thus approximately Gaussian distributed with mean $t_f$ and covariance $(\nu \t \hat{f}_f)^{-2} \nu \t P(t_f) \nu$.

Next, we substitute (\ref{eq:stopping_time_approx}) into the state approximation (\ref{eq:approximate_state_at_stopping_time}) to obtain an approximate expression of the state at the stopping time as
\begin{equation}
	x(T) \approx x(t_f) - \bigg( \frac{\hat{f}_f \nu \t}{\nu \t \hat{f}_f} \bigg) \tilde{x}(t_f) = \bar{x}(t_f) + \bigg( I - \frac{\hat{f}_f \nu \t}{\nu \t \hat{f}_f} \bigg) \tilde{x}(t_f)
\end{equation}
Since this approximation is a linear function of the nominal final state and the final state deviation, it follows that $\E \big( x(T) \big) \approx \bar{x}(t_f)$ and
\begin{equation} \label{eq:stae_cov_at_stopping_time}
	\Cov \big( x(T) \big) \approx Z_f P(t_f) Z_f \t, \quad Z_f = I - \frac{\hat{f}_f \nu \t}{\nu \t \hat{f}_f}
\end{equation}
\subsection{Stochastic Range Control with a State Trigger}


When using a state trigger, the entry guidance should aim to minimize the state error at the trigger time rather than at the final time.
The final state penalty in the range control cost (\ref{eq:long_range_cost}) is, accordingly, modified to
\begin{equation}
	\Var \, (a_{f,T} \t \tilde{x}_\ell(T)) \approx \Var \, (a_{f,T} \t Z_{\ell, f} \tilde{x}_\ell(t_f))
\end{equation}
\begin{equation}\label{eq:stopping_time_transform_matrix_long}
	Z_{\ell, f} = I - \frac{\hat{f}_{\ell, f} \nu \t}{\nu \t \hat{f}_{\ell, f}}, \quad \hat{f}_{\ell, f} =
	\begin{bmatrix}
		\hat{f}_{z_\ell} \\
		\hat{f}_\rho \dot{\hat{s}}
	\end{bmatrix}
\end{equation}
where $\hat{f}_{z_\ell}$ and $\hat{f}_\rho$ are the longitudinal vehicle and density drift terms evaluated at the nominal final states.
It follows that the effect of the state trigger can be included by setting the final state error weight $a_f$ in (\ref{eq:long_range_cost}) as a transformation of the desired weight $a_{f, T}$ on the states at the stopping time:
\begin{equation} \label{eq:final_weight_transform_stopping_time}
	a_f = Z_{\ell, f} \t a_{f, T}
\end{equation}


\section{Numerical Example}
\label{sec:numerical_examples}


In this section, we apply both the proposed stochastic range control and the Apollo final phase range control methods for a simulated entry problem at Mars using bank angle control.
Separate optimizations and Monte Carlo simulations are performed for both a fixed final time case and a velocity trigger case (as was done for MSL).
Further comparison of different state trigger criteria is left for future work.
For the simulation, Mars is assumed to be a sphere of uniformly distributed mass with gravitational parameter $\mu = \num{4.2828e13}$ \si{m^3/s^2}, rotating at a fixed rotation rate of $\Omega = \num{7.0882e-05}$ \si{rad/s}.
The vehicle properties, which are based on MSL, are listed in Table~\ref{table:msl_vehicle} and the nominal entry conditions are listed in Table~\ref{table:init_state}.
The simulation initializes at entry interface (EI), which is the point where the vehicle radius passes through the radius of the edge of the atmosphere $h_{\mathrm{atm}} = 125$ \si{km}, and terminates at the end of heading alignment when the planet-relative velocity drops below 500 \si{m/s}.
In the fixed final time scenario, the simulation terminates at the time when the nominal trajectory reaches 500 \si{m/s}.
Sources of uncertainty included in the Monte Carlo simulation are listed in Table~\ref{table:mc_uncert}; the initial state dispersions given in Table~\ref{table:init_state} are conservative for illustrative purposes.
The heading alignment phase begins at 1.1 \si{km/s}, during which the bank angle commands are given by the heading alignment law (\ref{eq:ha_law}) with the gain $K_{\mathrm{ha}} = 50$.
Bank angles commanded during the heading alignment phase are saturated to be within $\pm 45 ^ \circ$ for velocities between 1.1 \si{km/s} and 0.9 \si{km/s}, and within $\pm 35 ^ \circ$ otherwise.
We do not consider navigation uncertainty as our focus is guidance performance, and thus the controller has access to the exact vehicle state and atmospheric density for the purposes of feedback.

The target landing site is at the surface radius with position coordinates
\begin{equation}
	\theta_{\mathrm{targ}} = 137 ^ \circ, \quad \phi_{\mathrm{targ}} = 0 ^ \circ
\end{equation}
and the nominal vehicle position at EI was set to
\begin{equation}
	\hat{\theta}_{0} = 125.973 ^ \circ, \quad \hat{\phi}_{0} = 0 ^ \circ
\end{equation}
so that the simulation would nominally end with the vehicle 10 \si{km} uprange of the target point when the other initial vehicle states are as in Table~\ref{table:init_state}.

\begin{figure}
	\centering
	\tikzsetnextfilename{mars_delrho}
	\begin{tikzpicture}
	\begin{axis}[
		xlabel={Density variation (\%)},
		ylabel={Altitude (\si{km})},
		ymin=0, ymax=125,
		ytick={0,25,50,75,100,125},
		xtick={-75,-50,-25,0,25,50,75},
		grid = major,
		grid style = {draw=black!8},
	 	ylabel near ticks,
    	ylabel shift = -3 pt,
	]
		\pgfplotstableread{data/mars_delrho.txt}\loadedtable
		
		\addplot [black!40, x filter/.expression={x*100},] table [y=h, x=s1] {\loadedtable};
		\addplot [black!80, x filter/.expression={x*100},] table [y=h, x=s6] {\loadedtable};
		
		\addplot [dashed,dash pattern=on 5pt,black!100, x filter/.expression={+x*200},] table [x index=1, y index=0] {data/mars_atmo_stddev.txt};
		\addplot [dashed,dash pattern=on 5pt,black!100, x filter/.expression={-x*200},] table [x index=1, y index=0] {data/mars_atmo_stddev.txt};
		
		\node[anchor=north east] (source) at (axis cs:-50, 70){$2\sigma$};
		\draw[-stealth] (axis cs:-55, 68) to[bend right=-10] (axis cs:-35, 80);
		
		\node[anchor=north] (source) at (axis cs:48, 70){Samples};
		\draw[-stealth] (axis cs:48, 70) to[bend left=-10] (axis cs:35, 90);
		
	\end{axis}
\end{tikzpicture}
	\caption{
		MarsGRAM density variation samples
		\label{fig:mars_delrho}}
\end{figure}
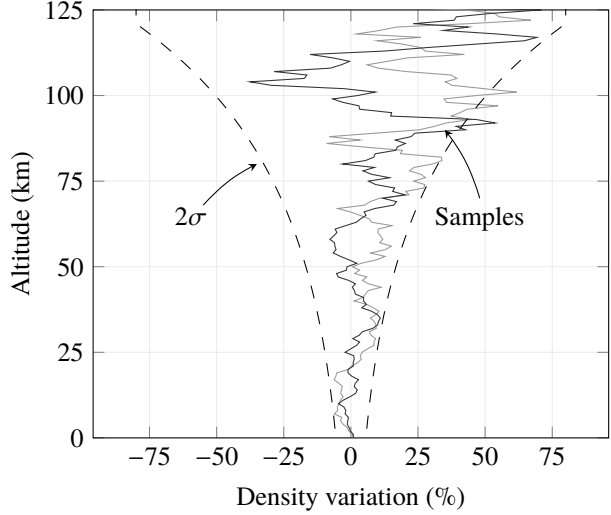

Both the nominal atmospheric density and the density variations are given as functions of altitude by MarsGRAM \cite{Justus2002}; resulting density variation samples are shown in Figure~\ref{fig:mars_delrho}.
The assumed density variation process is defined by setting $\lambda(s) = 2 / H(s)$ and $\varphi(s)$ as in Remark~\ref{remark:stationary_delrho_variance} for a desired variance profile $\zeta_d(s)$ computed from the MarsGRAM samples.
Note that this density variation process model is only used for the purposes of designing the feedback gains, and that when simulating the entry dynamics only the MarsGRAM density samples are used.
Additionally, the trimmed angle of attack was uniformly dispersed on the interval $[-16.5 ^ \circ, -14.5 ^ \circ]$, which, in turn, affected the $L/D$ and the ballistic coefficient (approx. errors of ~$\pm0.15 $ and $\pm 0.02$ in $C_L$ and $C_D$.).

%
%
%
%
%
%
%
%
%
%
%
%
%
%
%
%
%

The bank angle commands are recomputed each second, and the bank angle remains constant until the next command is issued.
Increments to the bank angle are constrained by a $\mathcal{L} = 15$ \si{deg/s} bank rate limit, and thus the bank angle $\sigma$ to be flown is a function of the new bank angle command $\sigma_{\mathrm{cmd}}$, the previous bank angle command $\sigma_{\mathrm{cmd,prev}}$, the rate limit $\mathcal{L}$, and the elapsed time $\Delta t$ from the time when the previous bank command was issued:
\begin{equation}
	\sigma = \begin{cases}
		 \sigma_{\mathrm{cmd,prev}} + \mathcal{L} \Delta t & \text{if} \;\; \sigma_{\mathrm{cmd}} - \sigma_{\mathrm{cmd,prev}} \geq + \mathcal{L} \Delta t \\
		 \sigma_{\mathrm{cmd,prev}} - \mathcal{L} \Delta t & \text{if} \;\; \sigma_{\mathrm{cmd,prev}} - \sigma_{\mathrm{cmd}} \leq - \mathcal{L} \Delta t \\
		\sigma_{\mathrm{cmd}} & \text{otherwise}
	\end{cases}
\end{equation}
For simplicity, the bank acceleration is not limited.

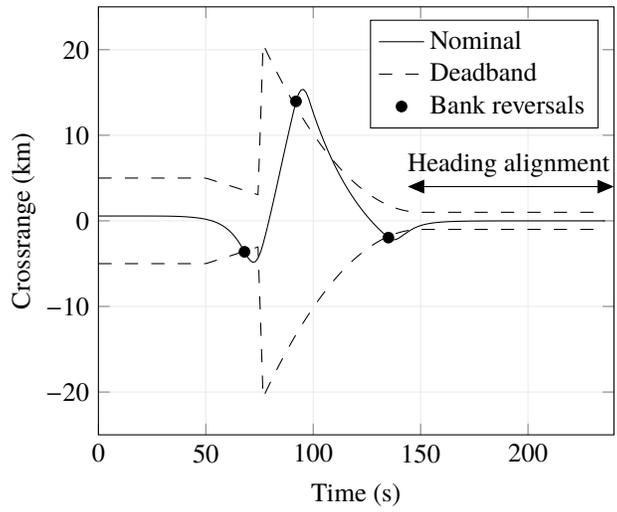
\begin{figure}
	\centering
	\tikzsetnextfilename{nom_crossrange}
		\begin{tikzpicture}

		\usetikzlibrary{arrows, arrows.meta}
		\begin{axis}[
		    xlabel={Time (\si{s})},
		    ylabel={Crossrange (\si{km})},
			xmin = 0, xmax=240,
		    ymin=-25, ymax=25,
		    legend pos=north east,
		    grid = major,
		    grid style = {draw=black!8},
	    	ylabel near ticks,
	    	ylabel shift = -3 pt,
	    	legend cell align=left,
		]
		
			\pgfplotsset{
			  /pgfplots/xlabel near ticks/.style={
			     /pgfplots/every axis x label/.style={
			        at={(ticklabel cs:0.5)},anchor=near ticklabel
			     }
			  },
			  /pgfplots/ylabel near ticks/.style={
			     /pgfplots/every axis y label/.style={
			        at={(ticklabel cs:0.5)},rotate=90,anchor=near ticklabel}
			     }
			  }
			  
			\addplot [black] table {data/mars_crossrange_nom.txt};
			\addplot
				[
					dashed,	
					dash pattern=on 5pt,
					y filter/.expression={-y},
				]
				table {data/mars_crossrange_deadband.txt};
			\addplot
				[
					dashed,	
					dash pattern=on 5pt,
					forget plot,
				]
				table {data/mars_crossrange_deadband.txt};
			\addplot [only marks,] table {data/mars_bank_rev_nom.txt};
		    \legend{Nominal, Deadband, Bank reversals,};
		
		    
		    \draw[>=triangle 45, <->] (axis cs:144, 4) to (axis cs:240, 4);
		    \node[anchor=south] (source) at (axis cs:191, 4){Heading alignment};
		\end{axis}
	\end{tikzpicture}
	\caption{
		Nominal crossrange trajectory with bank reversals
		\label{fig:crossrange_nominal}}
\end{figure}

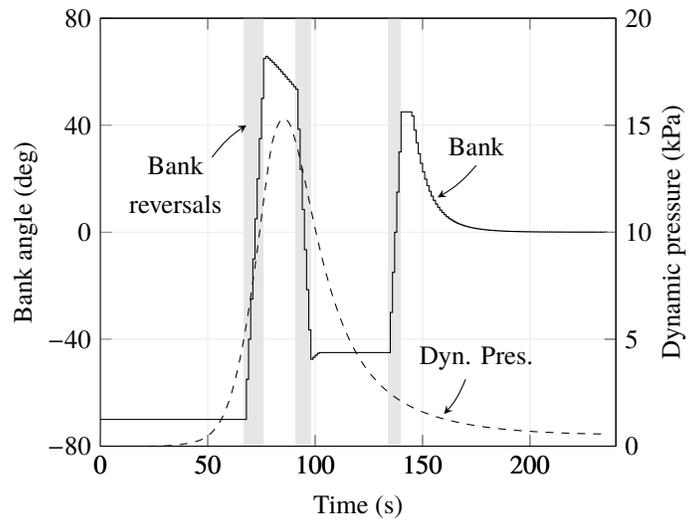
\begin{figure}
	\centering
	\tikzsetnextfilename{dyn_pres_bank}
	\begin{tikzpicture}
	\begin{axis}[
		ylabel = {Bank angle (\si{deg})},
		xlabel = {Time (\si{s})},
		ymin=-80, ymax=80,
		ytick={-80,-40,0,40,80},
		xmin=0, xmax=240,
		xtick={0,50,100,150,200},
		grid = major,
		grid style = {draw=black!8},
	 	ylabel near ticks,
    	ylabel shift = -3 pt,
	]
		\fill [black!10] (axis cs:67, -80) rectangle (axis cs:76, 80);
		\fill [black!10] (axis cs:91, -80) rectangle (axis cs:98, 80);
		\fill [black!10] (axis cs:134, -80) rectangle (axis cs:140, 80);
		
		\node[anchor=north] (source) at (axis cs:35, 35) {\begin{tabular}{c} Bank \\ reversals \end{tabular}};
		\draw[-stealth] (axis cs:52, 28) to[bend right=-10] (axis cs:68, 40);

		\node[anchor=south] (source) at (axis cs:175, 25) {Bank};
		\draw[-stealth] (axis cs:175, 25) to[bend right=-10] (axis cs:157, 13.02);
		
		\node[anchor=south] (source) at (axis cs:175, -55) {Dyn. Pres.};
		\draw[-stealth] (axis cs:170, -55) to[bend right=-10] (axis cs:160, -68);

		\addplot [black,const plot,] table {data/mars_bank_cmd_nom.txt};
	\end{axis}
	\begin{axis}[
		ylabel = {Dynamic pressure (\si{kPa})},
		ymin=0, ymax=20,
		xmin=0, xmax=240,
		axis y line*=right,
	 	ylabel near ticks,
    	ylabel shift = -3 pt,
	]
		\addplot [black,dashed] table {data/mars_dyn_pres_nom.txt};
	\end{axis}
\end{tikzpicture}
	\caption{
		Timing of bank reversals with dynamic pressure
		\label{fig:dyn_pres_bank}}
\end{figure}

\begin{figure*}
	\centering
	\tikzsetnextfilename{mars_nominal_trajectory}
	\centerfloat
	\begin{tikzpicture}

	\pgfplotsset{yticklabel style={text width=1.4em,align=right}}

	\pgfplotsset{
	 /pgfplots/xlabel near ticks/.style={
	   /pgfplots/every axis x label/.style={
	      at={(ticklabel cs:0.5)},anchor=near ticklabel
	   }
	 },
	 /pgfplots/ylabel near ticks/.style={
	   /pgfplots/every axis y label/.style={
	      at={(ticklabel cs:0.5)},rotate=90,anchor=near ticklabel}
	   }
	 }
	 
	\pgfplotsset{
		compat=1.3,
		every axis/.append style={scale only axis,
		height=4.25cm, width=4.9cm,
		xmin=0, xmax=240,
		ymajorgrids=true,
		xmajorgrids=true,
		xtick={0,50,100,150,200},
		grid style = {draw=black!8},
		ylabel near ticks,
		xlabel near ticks,
		ylabel shift = -4 pt,
  		}
	}

	\pgfplotstableread{data/mars_nom_traj.txt}\loadedtable
	
 	\begin{axis}[%
		name = plot1,
		ylabel = {Altutude (\si{km})},
		ymin = 0,
 	]
   		\addplot [black] table [x=t, y=h] {\loadedtable};
 	\end{axis}

	\begin{axis}[%
		name = plot2,
		at = (plot1.right of south east), anchor=left of south west,
		xshift = 4pt,
		ylabel = {Velocity (\si{km/s})}
	]
    	\addplot [black] table [x=t,y=v] {\loadedtable};
 	\end{axis}

	 \begin{axis}[%
	  name = plot3,
	  at = (plot2.right of south east), anchor=left of south west,
	  ylabel = {Flight path angle (\si{deg})},
	  xshift = 4pt,
	 ]
    	\addplot [black] table [x=t, y=fpa] {\loadedtable};
	 \end{axis}

	 \begin{axis}[%
	  name=plot4,
	  at=(plot1.below south west), anchor=above north west,
  	    ylabel = {Density (\si{kg/m^3} $\times$ $10^3$)},
  	    xlabel = {Time (\si{s})},
  			ymin = 0,
  			y filter/.expression={y*1000},
 	]
    	\addplot [black] table [x=t, y=rho] {\loadedtable};
 	\end{axis}
 
	\begin{axis}[%
    name=plot5,
		at = (plot4.right of south east), anchor=left of south west,
	    ylabel = {Dynamic pressure (\si{kPa})},
	    xlabel = {Time (\si{s})},
  			ymin = 0,
  			xshift = 4pt,
   ]
   		\addplot [black] table [x=t, y=q] {\loadedtable};
   \end{axis}
   
   \begin{axis}[%
	    name=plot6,
		at = (plot5.right of south east), anchor=left of south west,
		ylabel = {Range traveled (\si{km})},
		xlabel = {Time (\si{s})},
		ymin = 0,
		xshift = 4pt,
    ]
		\addplot [black,] table {data/mars_rng_nom.txt};
    \end{axis}
    
\end{tikzpicture}
	\caption{
		Nominal entry trajectory
		\label{fig:mars_nominal_trajectory}}
\end{figure*}
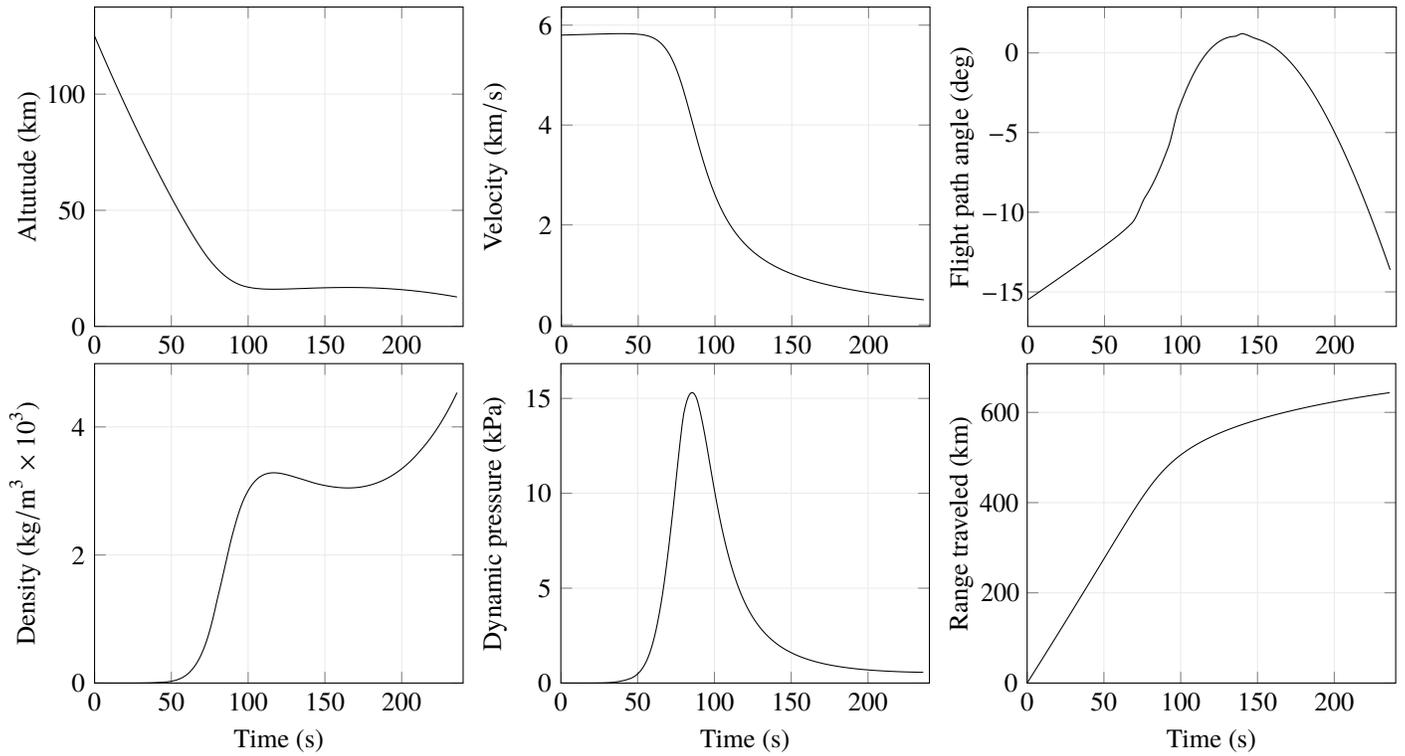

The nominal trajectory was obtained by integrating the equations of motion (\ref{eq:full_eom}) with the nominal density profile and the longitudinal control (cosine of the bank angle) given as a function of velocity.
In particular, the nominal bank angle cosine is equal to $\cos 75 ^ \circ$ for velocities above 5.5 \si{km/s}, $\cos 45 ^ \circ$ for velocities below 2.5 \si{km/s}, and a linear ramp from $\cos 75 ^ \circ$ to $\cos 45 ^ \circ$ for velocities between 5.5 \si{km/s} and 2.5 \si{km/s}.
The sign of each new bank angle command is determined by the deadband logic (\ref{eq:deadband_logic}) with the time-dependent deadband shown in Figure~\ref{fig:crossrange_nominal}.
The nominal bank angle profile, including bank reversals, is shown in Figure~\ref{fig:dyn_pres_bank}, and the resulting nominal entry trajectory is shown in Figure~\ref{fig:mars_nominal_trajectory}.

For both the Apollo and the stochastic range control methods, the nominal trajectory extends from EI to the end of the heading alignment phase, but the effect of control inputs is set to zero during the latter.
Considering the trajectory extended through the heading alignment phase allows for the range control guidance to target conditions at the end of heading alignment, rather than only being able to target conditions at the start of heading alignment.
The nominal longitudinal trajectories are obtained from the full nominal trajectory, but bank reversals are not included in the nominal control input; that is, the nominal longitudinal state $\hat{x}_\ell(t)$ at any time $t$ is computed as a function of the full nominal state $\hat{x}(t)$, whereas the nominal bank angle is only provided as the nominal bank angle cosine, as in the control law (\ref{eq:long_control_decomp}).
The Apollo final phase law is parameterized as a function of velocity, while the stochastic law is parameterized as a function of time.
The Apollo overcontrol gain is set to $K_\mathrm{oc} = 5$, and the targeting calculations aim for the target in inertial space at the current time (i.e., $t_{f, \mathrm{est}}$ in (\ref{eq:target_vector}) is set to equal $t$).

\begin{table}[]
	 \centering
	 \caption{Vehicle properties \label{table:msl_vehicle}}
	 \begin{tabular}[t]{lclp{1cm}lcl}
		 \hline \hline
		 Property & Value & Unit \\
		 \hline
		 Mass & 3200 & \si{kg} \\
 		 Reference area & $\pi 4.5 ^ 2 / 4$ & \si{m^2} \\
		 Trimmed angle of attack & -15.5 & \si{deg} \\
		 L/D & 0.24 &  \\
		 Ballistic coefficient & 135 & \si{kg/m^2} \\
		 \hline \hline
	 \end{tabular}
\end{table}

\begin{table}[]
	 \centering
	 \caption{Initial states \label{table:init_state}}
	 \begin{tabular}[t]{lccp{1cm}lcl}
		 \hline \hline
		 State & Nominal & $3\sigma$ \\
		 \hline
		 Altitude & 125 \si{km} & 0 \\
		 Velocity & 5.8 \si{km/s} & 20 \si{m/s} \\
		 Flight path angle & $-15.5 ^ \circ$ & $0.5 ^ \circ$ \\ 
		 Heading azimuth & $90.05 ^ \circ$ & $0.01 ^ \circ$ \\
		 Downrange & 0 & 5 \si{km} \\
		 Crossrange & 0.57 \si{km} & 0.5 \si{km} \\
		 \hline \hline
	 \end{tabular}
\end{table}

\begin{table}
	\centering
	\caption{Source of uncertainty included in the Monte Carlo \label{table:mc_uncert}}
	\begin{tabular}{cc}
		\hline \hline
		Source of uncertainty & Method \\
		\hline
		Vehicle state at EI & \Centerstack{Sampled from Gaussian\\ (Table~\ref{table:init_state})} \\
		Atmospheric density & MarsGRAM \\
		Vehicle aerodynamics & \Centerstack{Uniform distributed trim-$\alpha$\\ ($\pm0.15 $, $\pm 0.02$ in $C_L$, $C_D$)}
		   \\
		\hline \hline
	\end{tabular}
\end{table}

\subsection{Fixed Final Time}

First, we consider the range control problem with the final time fixed to be the final time of the nominal trajectory.
For the stochastic range control, the weights in the cost (\ref{eq:long_range_cost}) are set as
\begin{equation} \label{eq:final_time_example_final_penality}
	a_f \t = \begin{bmatrix}
		0 & 0 & 0 & 10^6 & 0
	\end{bmatrix}, \qquad \mathcal{R}_k = 10^{-2}
\end{equation}
The 99.73\%-ile of the closed-loop controls are constrained to lie within either $[-1, +1]$ (so that $\cos\inv$ in (\ref{eq:bank_angle_cmd_cosinv}) is defined) or $\pm0.45$ from the nominal value, whichever results in a smaller deviation from the nominal.
The limit $\pm0.45$ was chosen by trial and error to balance range control against crossrange performance.
This results in the control deviation limit given by
\begin{equation}
	\Delta^u_k = \min \{0.45, ~1 - \cos \hat{\sigma}(t^p_k), ~\cos \hat{\sigma}(t^p_k) + 1\}
\end{equation}
with $p^u_k = 1 - 0.9973$ ($3\sigma$) for every step $k$.
The final altitude and flight path angles are constrained to lie within $\pm 2$ \si{km} and $\pm1.55 ^ \circ$ respectively with probability at least $1 - 0.9973$ ($3\sigma$), which results in the constraint parameters
\begin{alignat}{2}
	d_1 &= \begin{bmatrix}
		1 & 0 & 0 & 0 & 0
	\end{bmatrix}\t, \qquad & \Delta^x_1 &= 2 \, \si{km}, \\
	d_2 &= \begin{bmatrix}
		0 & 0 & 1 & 0 & 0
	\end{bmatrix}\t, \qquad & \Delta^x_2 &= 1.55 ^ \circ
\end{alignat}
and $p^x_1 = p^x_2 = 1 - 0.9973$.
Per Theorem~\ref{thm:lq_equiv}, the values for $\Qlq$ and $\Rlq$ were found by minimizing the objective (\ref{eq:long_range_cost}), while satisfying the control and final state constraints, using MATLAB's \textsf{fmincon} function.

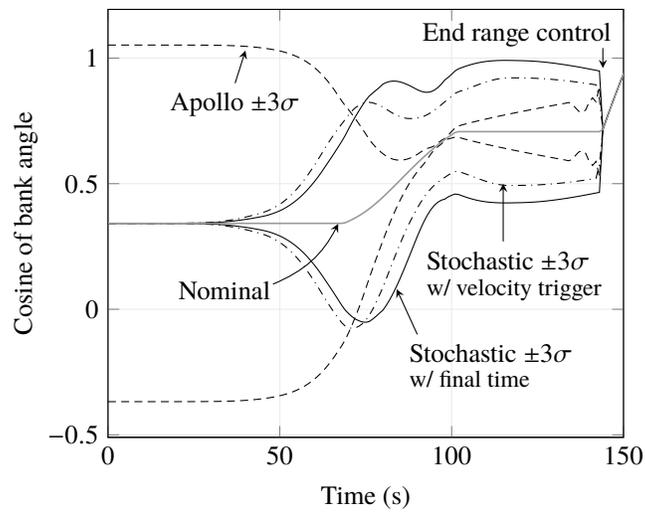
\begin{figure}
	\centering
	\tikzsetnextfilename{cosbank_pred_cov}
	\begin{tikzpicture}

\usepgfplotslibrary{fillbetween}
\usetikzlibrary{patterns}

\begin{axis}[
	ymin=-80, ymax=80,
	ytick=\empty,
	xmin=0, xmax=150,
	xtick=\empty,
		]
\end{axis}

\begin{axis}[
	ylabel = {Cosine of bank angle},
	xlabel = {Time (\si{s})},
	xmin=0, xmax=150,
	xtick={0,50,100,150,200},
	grid = major,
	grid style = {draw=black!8},
	ylabel near ticks,
	ylabel shift = -3 pt,
	]
	
%
%
	
	\addplot [black!100, densely dashed, thin, name path=apm,] table {data/lincov/cosbank_m3s_ap1.txt};
	\addplot [black!100, densely dashed, thin, name path=app,forget plot] table {data/lincov/cosbank_p3s_ap1.txt};

	\addplot [black!100, thin, name path=lqm,] table {data/lincov/cosbank_m3s_lq1.txt};
	\addplot [black!100, thin, name path=lqp,forget plot] table {data/lincov/cosbank_p3s_lq1.txt};
	
	\addplot [black!100, dashdotted, thin, name path=lqm,] table {data/lincov/cosbank_m3s_lq2.txt};
	\addplot [black!100, dashdotted, thin, name path=lqp,forget plot] table {data/lincov/cosbank_p3s_lq2.txt};
		

	

	\addplot [gray!80,semithick,] table {data/cosbank_nom_wo_bank_rev.txt};
	
	\node[anchor=north] (source) at (axis cs:37, 0.9) {Apollo $\pm3\sigma$};
	\draw[-stealth] (axis cs:37,0.9) to[] (axis cs:40, 1.03);
	
	\node[anchor=north west, align = left,execute at begin node=\setlength{\baselineskip}{0.95em},] (source) at (axis cs:85, -0.1) {Stochastic $\pm3\sigma$\\\small{w/ final time}};
	\draw[-stealth] (axis cs:90, -0.1) to (axis cs:84, 0.08);
	
	\node[anchor=south] (source) at (axis cs:119, 1.01) {End range control};
	\draw[-stealth] (axis cs:144, 1.04) to[] (axis cs:144, 0.96);
		
	\node[anchor=south east] (source) at (axis cs:50, 0) {Nominal};
	\draw[-stealth] (axis cs:48, 0.1) to[bend left=-20] (axis cs:67, 0.34);
		
	\node[anchor=south west, align=left,execute at begin node=\setlength{\baselineskip}{0.95em},] (source) at (axis cs:90, 0) {Stochastic $\pm 3\sigma$\\\small{w/ velocity trigger}};
	\draw[-stealth] (axis cs:115, 0.27) to[] (axis cs:115, 0.50);
	
\end{axis}
\end{tikzpicture}
	\caption{
		Longitudinal control input variance during range control computed via the linear covariance approximation
		\label{fig:cosbank_pred_cov}}
\end{figure}

Performance of both the Apollo and the stochastic entry guidance methods were evaluated by the linear covariance (LC) approximation as in (\ref{eq:long_cov_dynamics}) and by Monte Carlo simulation. 
The control input variances, based on the LC model, are shown in Figure~\ref{fig:cosbank_pred_cov}; while the input variance is different for the final time and the velocity trigger scenarios, both Apollo final phase solutions are similar and so only the final time solution is plotted.
Note that, per the control chance constraint (\ref{eq:long_control_constraints}), the $\pm3\sigma$ bounds for stochastic controllers remain within $[-1, +1]$, whereas the bounds for the Apollo controller at times leave this interval (which, in turn, degrades the LC approximation as it does not include saturation).
Interestingly, since the stochastic controller is derived through an optimization which takes dynamic pressure into account, the stochastic controller does not command bank angle corrections immediately following EI when the dynamic pressure is small.
Instead, the stochastic controllers both command more aggressive corrections than the Apollo law following the increase in dynamic pressure.

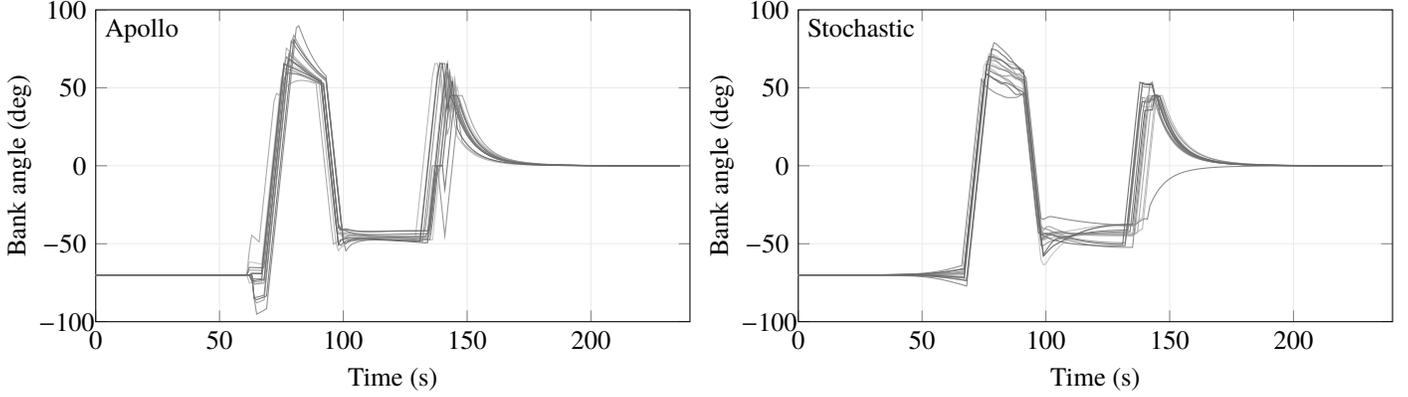
\begin{figure*}
	\centering
	\centerfloat
	\tikzsetnextfilename{mars_mc_control}
		\begin{tikzpicture}

		\pgfplotsset{yticklabel style={text width=1.8em,align=right}}

		\pgfplotsset{
		  /pgfplots/xlabel near ticks/.style={
		     /pgfplots/every axis x label/.style={
		        at={(ticklabel cs:0.5)},anchor=near ticklabel
		     }
		  },
		  /pgfplots/ylabel near ticks/.style={
		     /pgfplots/every axis y label/.style={
		        at={(ticklabel cs:0.5)},rotate=90,anchor=near ticklabel}
		     }
		  }
		  
		\pgfplotsset{
				compat=1.3,
				every axis/.append style={scale only axis,
				height=4.15cm, width=7.9cm,
				xmin=0, xmax=240,
				ymajorgrids=true,
				xmajorgrids=true,
				xtick={0,50,100,150,200},
				grid style = {draw=black!8},
			    ylabel near ticks,
			    xlabel near ticks,
			    ylabel shift = -4 pt,
		    	lcVarStyle/.style={dashed, dash pattern=on 5pt, thick},
		    	mcVarStyle/.style={dotted, thick},
		    	deadbandStyle/.style={dashed, dash pattern=on 5pt},
		    	sampleStyle/.style={black!50},
		    },
		}
		
		\def\numplot{15}
		\def\yminBank{-100}
		\def\ymaxBank{+100}
		\def\stochasticName{Stochastic}
		\def\colorScaleRange{40}
		\def\colorRangeMin{20}
		
		
		
		
	    \begin{axis}[%
			name = plot1,
			ylabel = {Bank angle (\si{deg})},
			xlabel = {Time (\si{s})},
			ymin = \yminBank,
			ymax = \ymaxBank,
	    ]
   			
	    	\foreach [evaluate=\m as \colorscale using (\m/\numplot)*\colorScaleRange+\colorRangeMin] \m in {1,2,...,\numplot}{
	    	    \edef\temp{\noexpand\addplot[black!\colorscale] table [x=t, y = mc\m] {data/mc/mc_ap1_bank.txt};}
	    	    \temp
	    	}
	    	\node[anchor=north west] (source) at (rel axis cs:0, 1) {Apollo};
	    \end{axis}
	
	    \begin{axis}[%
			name = plot2,
			at = (plot1.right of south east), anchor=left of south west,
			xshift = 4pt,
			ylabel = {Bank angle (\si{deg})},
			xlabel = {Time (\si{s})},
			ymin = \yminBank,
			ymax = \ymaxBank,
		]
    		\foreach [evaluate=\m as \colorscale using (\m/\numplot)*\colorScaleRange+\colorRangeMin] \m in {1,2,...,\numplot}{
				\edef\temp{\noexpand\addplot[black!\colorscale] table [x=t, y = mc\m] {data/mc/mc_lq1_bank.txt};}
				\temp
	    	}
	    	\node[anchor=north west] (source) at (rel axis cs:0, 1) {\stochasticName};
	    \end{axis}

	\end{tikzpicture}
	\caption{
		Monte Carlo control trajectories with a fixed final time
		\label{fig:mars_mc_control}}
\end{figure*}

Closed-loop sample trajectories with both $\pm3\sigma$ bounds computed from the samples and from the LC approximation are shown in Figure~\ref{fig:mars_mc_results}; note that while only 20 sample trajectories are shown, the $3\sigma$ bounds are computed from all 1,000 sample trajectories.
Bank angle histories are shown in Figure~\ref{fig:mars_mc_control}, and final state errors are listed in Table~\ref{table:results}.
The stochastic controller makes aggressive corrective controls following peak dynamic pressure which result in large altitude and flight path angle deviations, which is required in order to affect drag and thus range. 
Indeed, when using the stochastic controller, the range error begins to decrease following the increase in flight path angle error around 100 \si{s} after EI.
Furthermore, there is a trade-off between final altitude error and final range error, which is demonstrated by the final altitude with the stochastic controller error equaling the maximum allowed value of $\pm 2$ \si{km} with $3\sigma$ confidence.
While for most of the states plotted in Figure~\ref{fig:mars_mc_results} the LC $\pm 3\sigma$ approximation is close to the $\pm 3\sigma$ bounds computed from Monte Carlo, the LC approximation deviates most strongly from the Monte Carlo results for the flight path angle.
There are several possible causes for this approximation error: the exclusion of bank reversals from the LC approximation; the effects of closed-loop controls during the heading alignment phase, which are not included in the LC model; the LC approximation does not include the uniformly distributed error in the trimmed angle of attack; or statistical errors due to the finite number of Monte Carlo trials.

As one would expect, the increased intensity (i.e., the variance) of longitudinal control corrections with the stochastic controller comes with an increase in lateral dispersions.
The bank reversal logic together with the heading alignment phase successfully null out the additional crossrange error by the end of the heading alignment phase.


%
%

\begin{figure*}
	\centering
	\centerfloat
	\vspace{-0.5in}
	\tikzsetnextfilename{mars_mc_results}
	\input{tikzsrc/mars_mc_results.tikz}
	\caption{
		Monte Carlo trajectories with a fixed final time
		\label{fig:mars_mc_results}}
\end{figure*}

\subsection{Velocity Trigger}

Suppose now that the entry trajectories end when reaching a planet-relative velocity of 500 \si{m/s}.
This velocity trigger is described, as in (\ref{eq:stopping_time_def}), by the values
\begin{equation}
	\nu = \begin{bmatrix}
		0 & 1 & 0 & 0 & 0
	\end{bmatrix} \t, \qquad \beta = 500 \, \si{m/s}
\end{equation}
In units of \si{m}/\si{kg}/\si{s}, the stopping time transform matrix (\ref{eq:stopping_time_transform_matrix_long}) is given by
\begin{equation}
	Z_f = \begin{bmatrix}
		1 & -34.95 & 0  & 0 & 0 \\
		0 & 0 & 0  & 0 & 0 \\
		0 & -0.0015 & 1 & 0 & 0 \\
		0 & 144.75 & 0 & 1 & 0 \\
		0 & 1.5 \mathrm{e-}5 & 0 & 0 & 1
	\end{bmatrix}
\end{equation}
In order to minimize the range control cost (\ref{eq:long_range_cost}), with the same range-error penalty (\ref{eq:final_time_example_final_penality}) from the fixed time example, except applied to the state error at the time $T$ when the velocity reaches 500 \si{m/s}, we apply the transformation (\ref{eq:final_weight_transform_stopping_time}) to obtain the new final state error weight
\begin{equation}
	a_f \t =  \begin{bmatrix}
		0 & 0 & 0 & 10^6 & 0
	\end{bmatrix} Z_{\ell, f}
\end{equation}
The Apollo final phase guidance can similarly be improved by applying the velocity-trigger transformation to the final range error weight, which is given by the final state adjoint value in (\ref{eq:apollo_total_range_error}).
We thus set
\begin{equation}
	\vartheta_f = \begin{bmatrix}
		0 & 0 & 0 & 1 & 0
	\end{bmatrix} Z_{\ell, f}
\end{equation}

Sample Monte Carlo trajectories which use the velocity trigger are shown in Figures~\ref{fig:R_vs_V} and \ref{fig:R_vs_h}, where the improvement in range targeting performance with the stochastic controller is apparent.
Final state statistics derived from the full Monte Carlo simulation are listed in Table~\ref{table:results}.
Using the velocity trigger decreased the range errors, in addition to the obvious reduction in velocity error, for both the Apollo and stochastic controllers.
Furthermore, when using the velocity trigger, the lower 1\%-ile and upper 99\%-ile final range errors were approximately halved when using the stochastic controller in compared to the Apollo controller.

\begin{figure}
	\centering
	\tikzsetnextfilename{final_R_V_err}
	\begin{tikzpicture}
	\begin{axis}[
			ylabel = {Final velocity error (\si{m/s})},
			xlabel = {Final downrange error (\si{km})},
			grid = major,
			grid style = {draw=black!8},
			ylabel near ticks,
			ylabel shift = -3 pt,
			legend columns = 2,
			legend pos=north west,
		]
		\addplot [black,forget plot] table [y expr=\thisrow{y}*1000] {data/lincov/tf/lin_cov_tf_R_V_ap1.txt};
		\addplot [black,forget plot,dashed] table [y expr=\thisrow{y}*1000] {data/lincov/tf/lin_cov_tf_R_V_ap2.txt};
		\addplot [black,forget plot] table [y expr=\thisrow{y}*1000] {data/lincov/tf/lin_cov_tf_R_V_lq1.txt};
		\addplot [black,forget plot,dashed] table [y expr=\thisrow{y}*1000] {data/lincov/tf/lin_cov_tf_R_V_lq2.txt};

		
		
		
		\node[anchor=south west, align=left] (source) at (axis cs:1, 53) {Apollo \small{w/ final time}};
		\draw[-stealth] (axis cs:4, 53) to[bend right=-10] (axis cs:2.96, 35);
		
		\node[anchor=north east, align=right,execute at begin node=\setlength{\baselineskip}{0.95em},] (source) at (axis cs:12.5, 50) {Apollo\\\small{w/ velocity}\\\small{trigger}};
		\draw[-stealth] (axis cs:8, 25) to[bend right=-10] (axis cs:5.87, 12);

		\node[anchor=north west, align=left,execute at begin node=\setlength{\baselineskip}{0.95em},] (source) at (axis cs:-12.5, -18) {Stochastic\\\small{w/ velocity}\\\small{trigger}};
		\draw[-stealth] (axis cs:-10, -20) to[bend right=-10] (axis cs:-5.7, 30.8);
		
		\node[anchor=north west, align=left,execute at begin node=\setlength{\baselineskip}{0.95em},] (source) at (axis cs:-10, -57) {Stochastic \small{w/ final time}};
		\draw[-stealth] (axis cs:0, -60) to[bend right=-10] (axis cs:1.4, -50);

	\end{axis}
\end{tikzpicture}
	\caption{
		Range-velocity covariance at the final time
		\label{fig:R_V_cov}}
\end{figure}
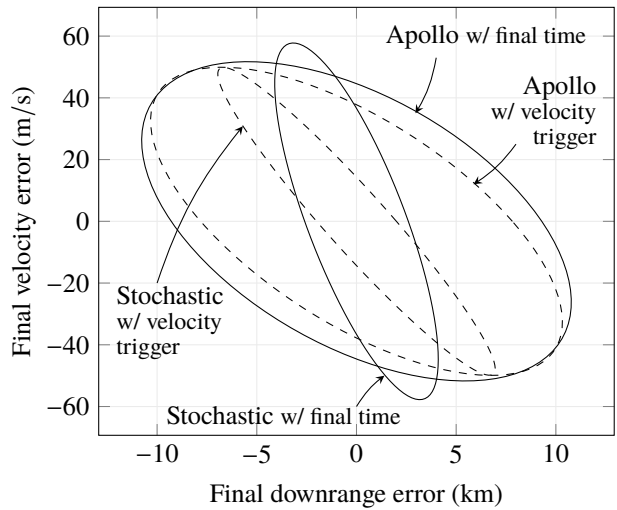

The range-velocity covariances at the final time, which are shown in Figure~\ref{fig:R_V_cov} (as computed by LC; ellipses contain $99.73\%$ probability), provide intuition behind the differences between the final time and the velocity trigger conditions.
While the Apollo final phase guidance naturally results in the range and velocity being negatively correlated --- with setting the final adjoint state per the velocity trigger further increasing this negative correlation --- there is a clear difference between the final time and velocity trigger stochastic guidance solutions.
For the final time case, the guidance is optimized to simply minimize the final range error, which is an objective not dependent on any final state correlations; thus, the range variance is decreased apparently at the expense of range-velocity correlation.
On the other hand, when including the effect of the velocity trigger in the optimization, the transformation (\ref{eq:final_weight_transform_stopping_time}) induces a penalty into the range cost, which depends on the final covariances.
Intuitively, we expect the velocity trigger cost to induce a strong correlation between velocity and range, since the conditional variance of range, provided the velocity will be fixed, approaches zero as the absolute value of the velocity-range correlation coefficient approaches one.
In other words, a strong correlation of two random variables implies that knowing the value of one variable strongly suggests the value of the other; thus, the variance for the final range error decreases when the final velocity is fixed per the trigger condition.

\begin{figure}
	\centering
	\tikzsetnextfilename{final_R_vs_V}
	\begin{tikzpicture}
\begin{groupplot}[
    group style={
        group name=my plots,
        group size=2 by 1,
		x descriptions at=edge bottom,
		y descriptions at=edge left,
        vertical sep=0pt,
		horizontal sep = 0pt,
    },
	ylabel={Velocity (\si{km/s})},
	xlabel={Range to go (\si{km})},
	xmin= -5, xmax = 5,
	ymin = 0.49, ymax = 0.55,
	grid = major,
	grid style = {draw=black!8},
	ylabel near ticks,
	ylabel shift = -3 pt,
	width = 2.1in,
	height = 3in,
	x dir = reverse,
]

\def\numplot{20}
\def\stochasticName{Stochastic}
\def\colorScaleRange{40}
\def\colorRangeMin{20}

\pgfplotsset{select coords between index/.style 2 args={
    x filter/.code={
        \ifnum\coordindex<#1\def\pgfmathresult{}\fi
        \ifnum\coordindex>#2\def\pgfmathresult{}\fi
    }
}}

\nextgroupplot

\foreach [evaluate=\m as \colorscale using (\m/\numplot)*\colorScaleRange+\colorRangeMin] \m in {1,2,...,\numplot}{
	\edef\temp{\noexpand\addplot[black!\colorscale] table [x=Rgo, y=V] {data/mc/vf/ap2mc_\m.txt};}
	\temp
}
\addplot [
	only marks, mark = *, mark size = 1.25, black,
	select coords between index={0}{\numplot}
	] table[x=Rgo, y=V] {data/mc/vf/ap2mc_final.txt};
\node[
	anchor=north east,
	align=right,
	execute at begin node=\setlength{\baselineskip}{0.93em},
	] (source) at (rel axis cs:1, 1) {Apollo\\\small{w/ velocity trigger}};
	
\nextgroupplot

\foreach [evaluate=\m as \colorscale using (\m/\numplot)*\colorScaleRange+\colorRangeMin] \m in {1,2,...,\numplot}{
	\edef\temp{\noexpand\addplot[black!\colorscale] table [x=Rgo, y=V] {data/mc/vf/lq2mc_\m.txt};}
	\temp
}
\addplot [
	only marks, mark = *, mark size = 1.25, black,
	select coords between index={0}{\numplot}
	] table[x=Rgo, y=V] {data/mc/vf/lq2mc_final.txt};
\node[
	anchor=north east,
	align=right,
	execute at begin node=\setlength{\baselineskip}{0.93em},
	] (source) at (rel axis cs:1, 1) {\stochasticName\\\small{w/ velocity trigger}};

\end{groupplot}

\end{tikzpicture}
	\caption{
		Sample trajectories terminating at the velocity trigger
		\label{fig:R_vs_V}}
\end{figure}
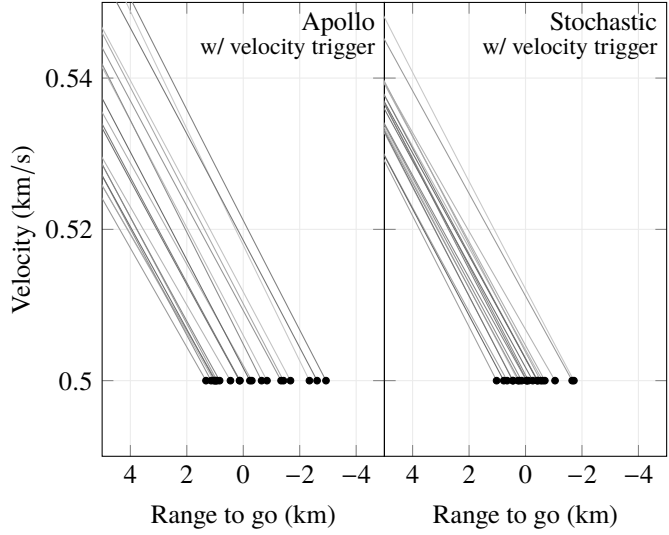

\begin{figure}
	\centering
	\tikzsetnextfilename{final_R_vs_h}
	\begin{tikzpicture}

\begin{groupplot}[
    group style={
        group name=my plots,
        group size=1 by 2,
		x descriptions at=edge bottom,
		y descriptions at=edge left,
        vertical sep=0pt,
    },
	ylabel={Altitude (\si{km})},
	xlabel={Range to go (\si{km})},
	xmin= -5, xmax = 5,
	ymax = 14.5,
	ytick = {11, 12, 13, 14},
	grid = major,
	axis equal,
	grid style = {draw=black!8},
	ylabel near ticks,
	ylabel shift = -3 pt,
	width = 3.75in,
	height = 1.82in,
	x dir = reverse,
]

\def\numplot{20}
\def\stochasticName{Stochastic}
\def\colorScaleRange{40}
\def\colorRangeMin{20}

\pgfplotsset{select coords between index/.style 2 args={
    x filter/.code={
        \ifnum\coordindex<#1\def\pgfmathresult{}\fi
        \ifnum\coordindex>#2\def\pgfmathresult{}\fi
    }
}}

\nextgroupplot

\foreach [evaluate=\m as \colorscale using (\m/\numplot)*\colorScaleRange+\colorRangeMin] \m in {1,2,...,\numplot}{
	\edef\temp{\noexpand\addplot[black!\colorscale] table [x=Rgo, y=h] {data/mc/vf/ap2mc_\m.txt};}
	\temp
}
\addplot [
	only marks, mark = *, mark size = 1.25, black,
	select coords between index={0}{\numplot}
	] table[x=Rgo, y=h] {data/mc/vf/ap2mc_final.txt};
\node[
	anchor=north east,
	align=right,
	execute at begin node=\setlength{\baselineskip}{0.93em},
	] (source) at (rel axis cs:1, 1) {Apollo\\\small{w/ velocity trigger}};

\nextgroupplot

\foreach [evaluate=\m as \colorscale using (\m/\numplot)*\colorScaleRange+\colorRangeMin] \m in {1,2,...,\numplot}{
	\edef\temp{\noexpand\addplot[black!\colorscale] table [x=Rgo, y=h] {data/mc/vf/lq2mc_\m.txt};}
	\temp
}
\addplot [
	only marks, mark = *, mark size = 1.25, black,
	select coords between index={0}{\numplot}
	] table[x=Rgo, y=h] {data/mc/vf/lq2mc_final.txt};
\node[
	anchor=north east,
	align=right,
	execute at begin node=\setlength{\baselineskip}{0.93em},
	] (source) at (rel axis cs:1, 1) {\stochasticName\\\small{w/ velocity trigger}};

\end{groupplot}

\end{tikzpicture}
	\caption{
		Sample trajectories with endpoints at the velocity trigger
		\label{fig:R_vs_h}}
\end{figure}
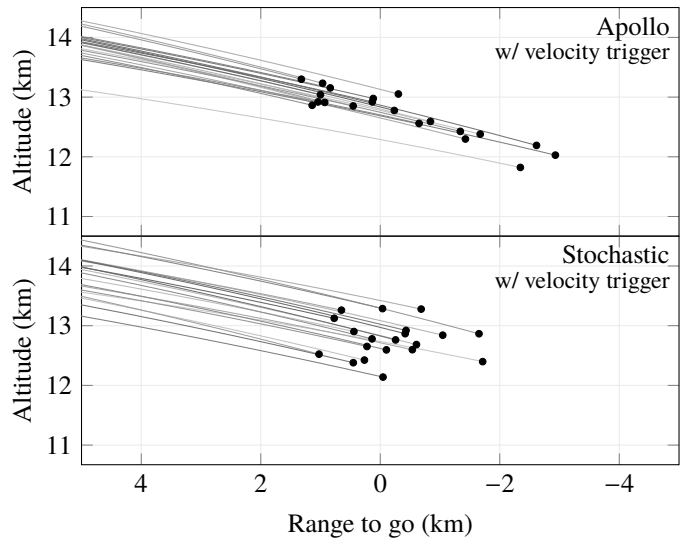

\begin{table*}
	\centering
	\caption{Final state error 1\textsuperscript{st} and 99\textsuperscript{th} percentiles from 1,000 Monte Carlo trials \label{table:results}}
	\begin{tabular}{lcccc}
		\hline \hline
		& \multicolumn{2}{c}{Apollo} & \multicolumn{2}{c}{Stochastic} \\
		\multicolumn{1}{r}{Trigger type:} & Time & Velocity & Time & Velocity \\
		\hline
		Downrange (\si{km}) & -8.36 / 6.53 & -3.58 / 3.04 & -4.64 / 2.56 & -1.96 / 1.19\\
Altitude (\si{km}) & -1.59 / 1.99 & -0.92 / 0.92 & -1.74 / 1.79 & -0.87 / 0.82\\
Velocity (\si{m/s}) & -39.54 / 42.89 & 0 / 0 & -54.59 / 51.97 & 0 / 0\\
Flight path angle (\si{deg}) & -1.28 / 1.55 & -2.37 / 2.13 & -1.46 / 2.15 & -2.07 / 2.05\\
		\hline \hline
	\end{tabular}
\end{table*}

\section{Conclusion}
\label{sec:conclusion}

In this paper, a stochastic process model for atmospheric entry in a randomly perturbed atmosphere was derived and then applied to develop a novel range control feedback law.
In contrast to the many existing models that include only parametric uncertainty, such as the scale height of an exponential atmosphere being random, the developed stochastic process model considers the atmospheric density uncertainty as a random function of altitude, similar to the GRAM dispersion models.
The effect of altitude-dependent density perturbations on the vehicle trajectory was obtained by setting the diffusion coefficient in the SDE model to be a function of the vehicle sink rate --- the faster the vehicle descends through altitude dependent perturbations, the more intense the time dependent perturbations. 
In future works, this model could enable rapid onboard uncertainty quantification to support decision making during entry.

The proposed guidance algorithm is implemented as a linear feedback law using table lookup, in the same manner as the flight proven Apollo final phase guidance algorithm; the difference being that the feedback gains in the proposed guidance law are derived from an optimization over the feedback gains with respect to the covariance evolution of the closed-loop system.
Furthermore, an analytical approximation for the effect of a state triggered termination of the entry trajectory was developed and applied to the proposed guidance algorithm.
In a Monte Carlo simulation of an MSL-like entry scenario at Mars, the proposed stochastic entry guidance results in the  1\%-ile and 99\%-ile of the final range errors being approximately halved when compared to the Apollo final phase guidance.

\appendix
\section*{Appendix}

\subsection{Proof of Lemma~\ref{lem:brownian_time_change}} \label{appendix:proof_brownian_time_change}

	The Dambis-Dubins-Schwarz theorem \cite[5.13]{LeGall2016} states that all continuous local martingales may be given as a time-changed Brownian motion: for a martingale $y(t)$ with $t \in [t_0, t_f]$, there exists a Brownian motion $\zeta$ such that, almost surely (a.s), for all $t \in [t_0, t_f]$, $y(t) = \zeta( \xi(t) )$, where $\xi(t)$ is the quadratic variation of $y$ \footnote{For a definition of quadratic variation see \cite[Theorem~4.9]{LeGall2016}.}.
	To show (\ref{eq:brownian_time_change_integral_form}), we let the martingale be given by
	\begin{equation}
		y(t) = \int_{t_0}^{t} \sqrt{ \dot{s}(a) } \, \d w(a)
	\end{equation}
	This martingale has quadratic variation $s(t) - s(t_0)$.
	Since Brownian motion is a stationary process, we can drop the constant term $s(t_0)$, and then we take the time changed Brownian motion $\tilde{w}( s(t) )$ to be $\zeta (s(t))$, which gives the desired result.

\subsection{Proof of Theorem~\ref{thm:lq_equiv}} \label{appendix:proof_lq_equiv}
	First, we rewrite the constraints (\ref{eq:long_final_state_constraints}) and (\ref{eq:long_control_constraints}) as limits on the covariance of the final state and the variance of the closed-loop control inputs.
	Since the inner product $d_i \t \tilde{x}_\ell(t_f) $ is a zero-mean Gaussian random variable with variance $d_i \t P_{\ell, N_p} d_i$, we have that
	\begin{equation}
		\Pr ( \vert d_i \t \tilde{x}_\ell(t_f) \vert \leq \Delta^x_i ) = \erf \bigg( \frac{\Delta^x_i}{\sqrt{2 d_i \t P_{\ell, N_p} d_i} } \bigg)
	\end{equation}
	in terms of the error function $\erf$.
	Rearranging, we obtain an equivalent constraint to (\ref{eq:long_final_state_constraints}), given by
	\begin{equation} \label{eq:long_final_state_constraint_rewritten}
		d_i \t P_{\ell, N_p} d_i \leq \Dscr^2_i, \quad \text{where} \quad \Dscr_i = \frac{\Delta^x_i}{\sqrt{2} \erf \inv (1 - p^x_i)}
	\end{equation}
	for $i = 1, \dots, N_x$.
	Similarly, since the control correction $\tilde{u}_{\ell, k}$ is zero-mean Gaussian distributed, we have that
	\begin{equation}
		\Pr ( \vert \tilde{u}_{\ell, k} \vert \leq \Delta^u_k ) = \erf \bigg( \frac{\Delta^u_k}{\sqrt{2 \Var \, (\tilde{u}_{\ell, k})}} \bigg)
	\end{equation}
	Substituting $\Var \, (\tilde{u}_{\ell, k}) = K_{\ell, k} P_{\ell, k} K_{\ell, k} \t$ and rearranging, we obtain an equivalent constraint to (\ref{eq:long_control_constraints}) given by
	\begin{equation} \label{eq:long_control_constraint_rewritten}
		K_{\ell, k} P_{\ell, k} K_{\ell, k} \t \leq \Uscr_k^2, \quad \text{where} \quad \Uscr_k = \frac{\Delta^u_k}{\sqrt{2} \erf \inv (1 - p^u_k)}
	\end{equation}
	for $k = 0, \dots, N_p - 1$.
	
	Next, we relate the LQ cost function (\ref{eq:std_lqg_cost}) to the Lagrangian for Problem~\ref{prob:stochastic_range_control}, which, with the Lagrange multipliers $\xi \in \Re ^ {N_x + N_p}$, is given as
	\begin{equation} \label{eq:long_lagrangian}
		L_\ell(K, \xi) = J_\ell(K) + \sum_{i = 1}^{N_x} \xi_i \big( d_i \t P_{\ell, N_p} d_i - \Dscr^2_i \big)
		+ \sum_{k = 0}^{N_p - 1} \xi_{N_x + k + 1} \big( K_{\ell, k} P_{\ell, k} K_{\ell, k} \t - \Uscr_k^2 \big)
	\end{equation}
	The terms in the Lagrangian that depend on the final state covariance are rearranged as
	\begin{multline} \label{eq:lagrangian_state_rearrange}
		\Var \, ( a_f \t \tilde{x}_\ell(t_f) ) + \sum_{i = 1}^{N_x}  \xi_i \big( d_i \t P_{\ell, N_p}  d_i - \Dscr_i^2 \big) \\
		= \tr \big( a_f a_f \t P_{\ell, N_p} \big) + \tr \bigg\{ \bigg( \sum_{i = 1}^{N_x} \xi_i d_i d_i \t \bigg) P_{\ell, N_p} \bigg\} - \sum_{i = 1}^{N_x} \xi_i \Dscr_i^2 \\
		= \tr \bigg\{ \bigg( a_f a_f \t + \sum_{i = 1}^{N_x} \xi_i d_i d_i \t \bigg) P_{\ell, N_p} \bigg\} - \sum_{i = 1}^{N_x} \xi_i \Dscr_i^2
	\end{multline}
	Similarly, the terms including the control variance $K_{\ell, k} P_{\ell, k} K_{\ell, k} \t$ are rearranged as
	\begin{multline} \label{eq:lagrangian_control_rearrange}
		\sum_{k = 0}^{N_p - 1} \mathcal{R}_k K_{\ell, k} P_{\ell, k} K_{\ell, k} \t + \sum_{k = 0}^{N_p - 1} \xi_{N_x + k + 1} \big( K_{\ell, k} P_{\ell, k} K_{\ell, k} \t - \Uscr_k^2 \big) \\
		= \sum_{k = 0}^{N_p - 1} \big( \mathcal{R}_k + \xi_{N_x + k + 1} \big) K_{\ell, k} P_{\ell, k} K_{\ell, k} \t - \sum_{k = 0}^{N_p - 1}  \xi_{N_x + k + 1}  \Uscr_k^2
	\end{multline}
	Let $(K^*, \xi^*)$ be a solution pair to Problem~\ref{prob:stochastic_range_control}, and let
	\begin{equation}
		\Qlq^* = a_f a_f \t + \sum_{i = 1}^{N_x} \xi_i^* d_i d_i \t, \quad \Rlq_k^* = \mathcal{R}_k + \xi^*_{N_x + k + 1}
	\end{equation}
	Substituting (\ref{eq:lagrangian_state_rearrange}) and (\ref{eq:lagrangian_control_rearrange}), we rewrite the Lagrangian (\ref{eq:long_lagrangian}) as
	\begin{equation} \label{eq:long_lagrangian_rearranged}
		L_\ell(K, \xi^*) = J_{LQ}(K; \Qlq^*, \Rlq_k^*) - Z \xi^*
	\end{equation}
	where
	\begin{equation}
		Z = \begin{bmatrix}
			\Dscr_1^2 & \cdots & \Dscr_{N_x}^2 & \Uscr_{0}^2 & \cdots & \Uscr_{N_p - 1}^2 
		\end{bmatrix}
	\end{equation}
	Since the term $Z \xi^*$ in (\ref{eq:long_lagrangian_rearranged}) does not depend on the decision variable $K$, minimizing the Lagrangian over the decision variable at the optimal value of the multipliers $\xi^*$ is equivalent to minimizing the LQ cost with the weights $\Qlq^*$ and $\Rlq^*$.
	Therefore, for the particular weight values $\Qlq^*$ and $\Rlq^*$, the solution to Problem~\ref{prob:lq} also solves Problem~\ref{prob:stochastic_range_control}.

\subsection{Proof of Corollary~\ref{cor:solve_via_QR}} \label{appendix:proof_solve_via_QR}
	Let $\mathscr{F} : (\Rlq, \Qlq) \mapsto K$ be the mapping from the LQ weights to the optimal LQ gains, which is obtained by evaluating the equations (\ref{eq:lq_solution}).
	By Theorem~\ref{thm:lq_equiv}, for particular LQ weights $\Qlq^*$ and $\Rlq^*$, the solution to Problem~\ref{prob:lq} also solves Problem~\ref{prob:stochastic_range_control}.
	It follows that the gain $K^*$ which solves Problem~\ref{prob:stochastic_range_control} is in the range of $\mathscr{F}$, and hence, if we replace the decision variable $K$ with $K = \mathscr{F}(\Rlq, \Qlq)$, then we can obtain the optimal gain as $K^* = \mathscr{F}(\Rlq^*, \Qlq^*)$.

\subsection{Targeting Calculations} \label{sec:targeting_calculations}


In this section, we review the targeting calculations to solve for downrange and crossrange distances, based on Ref.~\cite{Moseley1969apollo}.

First, consider the following reference frame definitions:
The planet-centered inertial frame $\I$ with basis $\{\hat{\imath}^\I, \hat{\jmath}^\I, \hat{k}^\I \}$; the planet-centered planet fixed frame $\F$ with basis $\{ \hat{\imath}^\F, \hat{\jmath}^\F, \hat{k}^\F \}$, which is defined relative to the planet-centered inertial frame by the planet rotation angle $\eta$ about the $\hat{k}^\I$ axis; and the vehicle rotating frame $\R$ (also known as the up-east-north frame), which is defined relative to the planet-centered planet fixed frame by two rotations for the vehicle longitude and latitude so that $\hat{\imath}^\R$ is aligned with the vehicle position $\vec{r}$.
We use in this section the notation that a subscript denotes reference frame, and a superscript outside of brackets denotes a vector being expressed in a coordinate system.

A vector in planet-centered, planet-fixed coordinates is transformed into planet-centered inertial coordinates by the matrix
\begin{equation}
	T^{\I \F}(\eta) = \begin{bmatrix}
		\cos \eta & -\sin \eta & 0 \\
		\sin \eta & \cos \eta & 0 \\
		0 & 0 & 1
	\end{bmatrix}
\end{equation}
Similarly, when the vehicle is at longitude $\theta$ and latitude $\phi$ (planet-fixed), then vectors in vehicle-rotating coordinates $\R$ are transformed to planet-centered, planet-fixed coordinates $\F$ by the matrix
\begin{equation}
	T^{\F \R}(\theta, \phi) = \begin{bmatrix}
		\cos \theta \cos \phi & - \sin \theta & -\cos \theta \sin \phi \\
		\cos \phi \sin \theta & \cos \theta & -\sin \theta \sin \phi \\
		\sin \phi & 0 & \cos \phi
	\end{bmatrix}
\end{equation}
The vehicle position in planet-centered inertial coordinates is therefore given by
\begin{equation}
	[ \vec{r}\, ]^\I = T^{\I  \F}(\eta) T^{\F  \R}(\theta, \phi) \begin{bmatrix}
		r \\ 0 \\ 0
	\end{bmatrix} = \begin{bmatrix}
		r \cos \phi \cos (\eta + \theta) \\
		r \cos \phi \sin (\eta + \theta) \\
		r \sin \phi
	\end{bmatrix}
\end{equation}
The vehicle velocity in the planet-fixed frame is given in vehicle rotating coordinates as
\begin{equation}
	[ \vec{V}_{\F} ]^{\R}
	= \begin{bmatrix}
		V \sin \gamma \\ V \cos \gamma \sin \psi \\ V \cos \gamma \cos \psi
	\end{bmatrix}
\end{equation}
and it follows that the planet-relative vehicle velocity is planet-centered inertial coordinates is obtained by
\begin{equation}
	[ \vec{V}_\F ]^\I = T^{\I \F}(\eta) T^{\F  \R}(\theta, \phi)[ \vec{V}_{\F} ]^{\R}.
\end{equation}
Finally, the vehicle velocity in the planet-centered inertial frame and in planet-centered inertial coordinates is
\begin{eqnarray}
	[ \vec{V}_\I ]^\I = [ \vec{V}_\F ]^\I + [ \vec{\Omega} ]^\I \times [ \vec{r}\, ]^\I
\end{eqnarray}
where $[ \vec{\Omega} ]^\I = [0 \; 0 \; \Omega] \t$ is the planet rotation vector in planet-centered inertial coordinates.

The target position is defined by a target longitude $\theta_\mathrm{targ}$ and target latitude $\phi_\mathrm{targ}$.
This position is given in planet-fixed inertial coordinates at an estimated final time $t_{f, \mathrm{est}}$ as
\begin{equation} \label{eq:target_vector}
	[\hat{r}_{\mathrm{targ}}]^\I = T^{\I \F}(\eta_0 + t_{f, \mathrm{est}} \Omega) \begin{bmatrix}
		\cos \phi_\mathrm{targ} \cos \theta_\mathrm{targ} \\
		\cos \phi_\mathrm{targ} \sin \theta_\mathrm{targ} \\
		\sin \phi_\mathrm{targ}
	\end{bmatrix}
\end{equation}
where $\eta_0$ is the planet rotation angle at the initial time.
The downrange angle is given as
\begin{equation}
	\delta_{\mathrm{go}} = \cos \inv \big( [\hat{r}_{\mathrm{targ}}]^\I \cdot [ \hat{r} ]^\I \big),
\end{equation}
where $[ \hat{r} ]^\I = [ \vec{r} ]^\I / r$, and the crossrange angle is
\begin{equation}
	\varepsilon = \frac{\pi}{2} - \cos \inv \big( [\hat{r}_{\mathrm{targ}}]^\I  \cdot [\hat{h}]^\I \big)
\end{equation}
where $[\hat{h}]^\I$ is a unit vector pointing towards the vehicle angular momentum vector in planet-centered inertial coordinates, which is given as $[\vec{h}]^\I = [ \vec{r} \, ]^\I \times [ \vec{V}_\I ]^\I$.
The downrange and crossrange distances are then obtained by multiplying the angles $\delta_{\mathrm{go}}$ and $\varepsilon$ by the planet radius, or more generally, by any reference radius.

\section*{Funding Sources}

This work was supported by NASA Space Technology Research Fellowship award 80NSSC17K0093.

\section*{Acknowledgments}

The authors would like to thank Professor Michael Damron for his help with Lemma~\ref{lem:brownian_time_change}.

\bibliography{jgcd21b}

\end{document}